\documentclass[reprint, superscriptaddress, aps, showkeys, pra, twocolumn]{revtex4-2}
\usepackage{graphicx} 
\usepackage{amsmath}
\usepackage{mathtools}
\usepackage{easybmat}
\usepackage{braket}
\usepackage{amsfonts}
\usepackage{xcolor}
\usepackage[colorlinks=true,citecolor=blue,urlcolor=blue, linkcolor=magenta]{hyperref}

\newcommand{\dbra}[1]{ \langle \langle #1 |}
 \newcommand{\pt}{\mathcal{PT}-}

\begin{document}

\title{Thermodynamics of an Open Two-Level $\mathcal{PT-}$Symmetric Quantum System}
\author{Baibhab Bose\textsuperscript{}}
\email{Corresponding author; baibhab.1@iitj.ac.in}
\author{Devvrat Tiwari\textsuperscript{}}
\email{devvrat.1@iitj.ac.in}
\author{Subhashish Banerjee\textsuperscript{}}
\email{subhashish@iitj.ac.in}
\affiliation{Indian Institute of Technology Jodhpur-342030, India\textsuperscript{}}
\date{\today}

\begin{abstract}
    For a subclass of a general two-level $\pt$symmetric Hamiltonian obeying an anti-commutation relation with its conjugate, a Hermitian basis is found that spans the bi-orthonormal energy eigenvectors. Using the modified projectors constructed from these eigenvectors, the generalized density matrix of the $\pt$symmetric evolution is calculated, and subsequently, ergotropy for a closed system is obtained. The $\pt$symmetric system, in an open system scenario, is studied to understand ergotropy under different regimes of non-Hermiticity of the Hamiltonian. The consistency of the three laws of thermodynamics for the $\pt$symmetric system in an open system scenario is also analyzed.
\end{abstract}
\keywords{Non-Hermitian quantum mechanics, open quantum systems, $\mathcal{PT-}$symmetry, quantum thermodynamics}
\maketitle

\section{Introduction}
Beyond the regime of Hermiticity, the theory of quantum mechanics has also been developed for a Hamiltonian symmetric in parity and time reversal ($\mathcal{PT}$) without losing the crucial constraint of real eigenvalues~\cite{Bender1, Bender2_1999, Bender2002}. This class of non-Hermitian Hamiltonians, having a broader symmetry, has been explored along different points of view in recent years~\cite{Bender_4_2003, Bender_5_2005, Bender_6_2007, Asok_das_2010, neutosci1_PhysRevD.89.125014, Nori_2019, Das_Bhasin_2025, Kumari_2022}. The necessary and sufficient conditions for non-Hermitian Hamiltonians to have real spectra and other mathematical intricacies were investigated in various works~\cite{Mostafazadeh_1_2002, Mostafazadeh_2_2002, Mostafazadeh_3_2002, Mostafazadeh_4_2004, Mostaf_5_doi:10.1142/S0219887810004816,Brody_2014}. 
Non-Hermitian and parity–time $\pt$symmetric quantum systems have attracted significant attention due to their ability to exhibit entirely real spectra and unconventional dynamical features despite the absence of Hermiticity. Foundational theoretical developments established the consistency of $\pt$symmetric and pseudo-Hermitian quantum mechanics and clarified their relation to modified inner-product structures and unitary evolution in an extended sense \cite{Brody2016, Mostafazadeh_2002, Bender_6_2007, Choutri2017}. These ideas have since been explored in quantum field-theoretic and relativistic settings \cite{Bender2012, Alexandre2018PRD,Bender2018fieldtheory}, as well as in quantum information and finite-dimensional systems \cite{CrokePT_information, Brody2013, Deffner2015,Ashida2017}. On the experimental and applied side, $\pt$symmetry has been realized across a variety of platforms, including optical, mechanical, and engineered systems, where phase transitions, exceptional points~\cite{exceptional0,exceptional1,exceptional2,exceptional3,exceptional4}, and anomalous dynamical effects have been observed \cite{Optics_Rter2010, Peng2014NatureComm, Zheng2013, Xiao2017, Gao:21}. In parallel, several works have investigated finite-dimensional quantum systems, quantum information aspects, and dynamical properties of $\pt$symmetric models \cite{Brachio1, Brachio2, Wen2019, Wang2020}. More recently, attention has turned to the role of environmental effects and open-system dynamics in $\pt$symmetric settings, including the formulation of consistent dynamical descriptions and the study of dissipative behavior \cite{PTopen1, PTopen2Brody_2021, Stenholm01042004, PTopen3}. $\mathcal{PT-}$symmetric systems in the context of quantum correlations~\cite{Naikoo_2021, Javid_SB_2019}, quantum walk~\cite{walk_Javid_2020, Badhani_2024}, and open quantum systems~\cite{Nori_2019, TommyOhlsson_densitymatrixforma_PhysRevA.103.022218} have been studied. However, a systematic thermodynamic framework that consistently integrates $\pt$symmetry with general non-Markovian open quantum dynamics, particularly in terms of energy exchange, ergotropy, and work extraction, remains relatively underexplored. In this work, we aim to address this gap by developing a thermodynamic description of $\pt$symmetric open quantum systems within a consistent dynamical framework.
The formalism developed in this work, combining $\pt$symmetry with open-system quantum thermodynamics, can be relevant across several areas where non-Hermitian effective descriptions and environmental interactions naturally arise.
In particular, $\pt$symmetric systems are experimentally realized in optical platforms with balanced gain and loss \cite{ElGanainy2018, Feng2017, Optics_Rter2010, Javid_2019, Agarwal_2012}. 
Effective non-Hermitian Hamiltonians appear in driven-dissipative and open quantum materials, mesoscopic and condensed matter systems \cite{Ashida2020, Bergholtz2021}. 
$\pt$symmetric and non-Hermitian effects have been implemented in topo-electrical circuits \cite{Helbig2020, Hofmann2020}.
Controlled dissipation and engineered gain/loss in cold-atom platforms provide a setting where $\pt$symmetric open dynamics can be explored \cite{Li2019, Lapp2019}. 
Non-Hermitian descriptions are often used in reaction kinetics, population dynamics, and biological transport processes \cite{Moiseyev2011, Rotter2009}, as well as in lasers \cite{PRAVEENA2023171260}.

Various other fields including  electronics~\cite{Electronics_PhysRevA.84.040101}, microwaves~\cite{microwaves_PhysRevLett.108.024101}, mechanics~\cite{mechanical_Bender2013}, acoustics~\cite{acoustic_Fleury2015}, atomic physics~\cite{atomic_1_Baker1984,atomic_2_PhysRevLett.117.123601,atomic_1_PhysRevLett.110.083604} and single-spin systems~\cite{singlespin_doi:10.1126/science.aaw8205}, among others have put non-Hermitian Hamiltonians with $\mathcal{PT-}$symmetry in perspective. 
The framework of non-Hermitian theories with $\mathcal{PT}$ symmetry has been used to study a number of other fundamentally significant issues, such as neutrino oscillations~\cite{neutosci_2_Ohlsson2016}, neutrino mass generation~\cite{neutmassgen_Alexandre2015}, light neutrino masses in Yukawa theory~\cite{Yukawa_Alexandre2017}, spontaneous symmetry breaking and the Goldstone theorem~\cite{Goldstone_PhysRevD.98.045001}, and the Brout–Englert–Higgs mechanism~\cite{Brout_higgs_PhysRevD.99.045006, brout_higgs_2_PhysRevD.99.075024}.
Overall, our formalism provides a unified thermodynamic framework to analyze energy exchange, dissipation, and work extraction in a two-level $\pt$symmetric open system, with direct relevance to experimentally accessible platforms such as photonics, circuits, and cold atoms. 

A linear positive-definite operator $\eta$ constructed by the parameters of the non-Hermitian Hamiltonian characterizes the inner-product space containing the corresponding dynamics, such that the Hermiticity is supplanted by a more general class of $\eta$-pseudo Hermiticity, $\eta^{-1}H^{\dagger}\eta=H$, maintaining the energy eigenvalues real.
The $\eta$-pseudo Hermitian framework built by distinct right and left eigenvectors of the $\pt$symmetric Hamiltonian redefines the projectors and consequently the operators of the inner-product space.
An alternate formulation of this framework that accommodates the dynamics of a $\pt$symmetric Hamiltonian is done by finding a Hermitian basis that spans the bi-orthonormal eigenspace consisting of the right and left eigenvectors. In~\cite{Das_Bhasin_2025}, it is seen that for a particular subclass of a general $\mathcal{PT}$-symmetric Hamiltonian, obeying an anti-commutator algebraic structure permits the construction of a Hermitian basis, wherein the Hamiltonian and its complex-conjugate counterpart act as mutually connected ladder operators that interconnect the basis states. The eigenvectors of an appropriate Hermitian operator generate the Hermitian basis states, omitting the need to obtain the metric $\eta$. The Hermitian basis for this specific subclass of $\pt$symmetric Hamiltonians also illustrates the geometry of the right and left eigenvectors in the state space. Thus, any corresponding quantum mechanical observable can be determined in terms of the modified eigenspace for this specific subclass of $\pt$symmetric Hamiltonians.
In this paper, we are interested in calculating quantum thermodynamical quantities as well as the applicability of the laws of thermodynamics in the $\pt$symmetric setup. To gain a comprehensive understanding of the thermodynamic properties of a system, it is essential to consider its environment. For this reason, we apply the tools of open quantum systems, which analyze a quantum system coupled to its surrounding environment~\cite{Weiss2011, Breuer2007, Banerjee2018, Omkar2016, Vacchini_2011, Tiwari_2023, tiwari2024strong}. A well-established canonical approach to open systems is the Gorini-Kossakowski-Sudarshan-Lindblad (GKSL) type interaction modeling Markovian evolution~\cite{GKLSpaper, Lindblad1976}. Beyond the GKSL type interaction, non-Markovian evolution has been extensively studied in recent years~\cite{Hall_2014, Rivas_2014, RevModPhys.88.021002, CHRUSCINSKI20221, banerjeepetrucione, vega_alonso, Utagi2020, kading2025}.

The study of the dynamics of thermodynamic quantities in the presence of strong system-bath coupling in the non-Markovian regime is a challenging task~\cite{landi_RevModPhys.93.035008,Strasberg_2016, Zhang_2022, 7t26_13_PhysRevE.97.062108, tiwari2024strong}.
One of the most fundamental criteria of non-Markovian quantum thermodynamics is to have consistent definitions of basic thermodynamic quantities such as heat, work, internal energy, ergotropy, and entropy production ~\cite{subotnik_2018, Strasbegr_2019, Rivas_strong_coupling, full_counting_paper, Zhang_2021, Strasberg_Esposito_2017} and fluctuation theorems~\cite{Fluctu1, Fluctu2} in the quantum regime. This is essential for defining the thermodynamic laws of non-Markovian quantum systems. In our work, we couple a single-mode Hermitian radiation bath to the $\pt$symmetric system by a Jaynes-Cummings-type interaction and by tracing out the bath degrees of freedom from the total evolution, to obtain the generalized density matrix of the evolution, which would be non-Markovian in general~\cite{bose2025PT}. Using the generalized density matrix, quantum thermodynamic quantities such as ergotropy, heat, internal energy, work, entropy production, and von-Neumann entropy are calculated, and the consistency of the three laws of quantum thermodynamics is verified.

In this paper, the Hermitian basis for the non-Hermitian $\pt$symmetric Hamiltonian is discussed in Sec.~\ref{SecII} along with its exceptional and normal points of the parameter space. In Sec.~\ref{SecIII}, the analytic calculation of ergotropy is shown for a closed $\pt$symmetric system. Sec.~\ref{SecIV} shows the open system treatment of the $\pt$symmetric Hamiltonian and its ergotropy. Sec.~\ref{SecV} studies the laws of quantum thermodynamics for the $\pt$symmetric open system, followed by the conclusions in Sec.~\ref{SecVI}.

\section{Hermitian basis for a non-Hermitian Hamiltonian}\label{SecII}
The general form of a non-Hermitian and $\mathcal{PT-}$symmetric Hamiltonian whose eigenvectors are orthogonal with respect to $\mathcal{CPT}$ inner product is \cite{Bender2002,Bender_2004}
\begin{align}\label{Hohl}
    \mathcal{H} = \begin{pmatrix}
    re^{i\psi} & s \\
    s & re^{-i\psi}
    \end{pmatrix},
\end{align}
where, $r,s,\psi \geq 0$. This Hamiltonian is also $\eta$-pseudo Hermitian with respect to a Hermitian metric $\eta$, $\eta^{-1}H^{\dagger}\eta=H$~\cite{bose2025PT}. In~\cite{SimilarityinPT} a larger class of similarity transformation is illustrated that goes beyond unitary or anti-unitary symmetry. Other works~\cite{Mostafazadeh_2_2002,Choutri2017,SimilarityAshida02072020, similarity3,similarity2Mostafazadeh2003,Scolarici2007,Scolarici2006} also elaborate on the pseudo-Hermiticity of non-Hermitian Hamiltonians and its connection to $\pt$symmetry. 
In \cite{Das_Bhasin_2025}, a recipe for constructing a Hermitian basis for a non-Hermitian Hamiltonian of a specific class was discussed. The specification is given by the equation
\begin{align}
    \{H,H^{\dagger}\}=d\boldsymbol{I},
\end{align}
where $d$ is one of the eigenvalues of the anticommutator of a general two-dimensional non-Hermitian Hamiltonian $H$ and its complex conjugate $H^{\dagger}$, and $\boldsymbol{I}$ is the identity matrix.
Re-scaling the Hamiltonian $H\rightarrow H/\sqrt{d}$, we get a simpler version of the above equation that provides the constant of motion of the given dynamics,
\begin{align}\label{anticomu}
    \{H,H^{\dagger}\}=\boldsymbol{I}.
\end{align}
This constraint equation makes possible a corresponding anticommutator algebra with $H$ and $H^\dagger$ as its ladder operators. The basis on which they do so is constructed by using the eigenvectors of a Hermitian operator 
\begin{align}
    F=H^\dagger H.
\end{align}
$F$ is a positive Hermitian operator that encodes both $H$ and $H^\dagger$ having dimension of energy squared, and it has no information about the phase of the complex energy. The eigenstates of this operator are naturally orthogonal, and their eigenvalues are bounded from both sides: $0 \le f \leq 1$.
In the eigenvalue equation
\begin{align}\label{feigenVeq}
    F\ket{f}=f\ket{f},
\end{align}
we manipulate the algebra of the anticommutator constraint equation in Eq.~\eqref{anticomu} to find the raising and lowering operations of $H$ and $H^\dagger$.
In our work the two-dimensional $\pt$symmetric Hamiltonian in Eq.~\eqref{Hohl} obeys the constraint Eq.~\eqref{anticomu} only when $\psi=\pi/2$ and we have
\begin{align}
    H = \begin{pmatrix}
    ir & s \\
    s & -ir
    \end{pmatrix}.
\end{align}
The anticommutator relation is normalized as prescribed in Eq.~\eqref{anticomu} by rescaling the Hamiltonian. The value of the anticommutator is
\begin{align}
    \{H,H^\dagger\}=2(r^2+s^2)\boldsymbol{I}.
\end{align}
The rescaled Hamiltonian is
\begin{align}\label{Hpt}
    H = \frac{1}{\sqrt{d}}\begin{pmatrix}
    ir & s \\
    s & -ir
    \end{pmatrix},
\end{align}
where $d=2(r^2+s^2)$. This rescaled Hamiltonian will now be used in the subsequent work. 
The corresponding $F$ operator is
\begin{align}
    F=H^\dagger H=\frac{1}{2(r^2+s^2)}\begin{pmatrix}
        r^2+s^2 & -2irs\\
        2irs & r^2+s^2
    \end{pmatrix},
\end{align}
whose eigenvalues are
\begin{align}\label{feigenVa}
    f&=\kappa(s+r)^2, \nonumber \\
    f'&=\kappa(s-r)^2,
\end{align}
where $\kappa=\frac{1}{d}=\frac{1}{2(r^2+s^2)}$. It is seen that $f'=1-f$. The corresponding normalized eigenvectors are,
\begin{align}\label{feigenve}
    \ket{f}=\frac{1}{\sqrt{2}}\begin{pmatrix}
        -i \\
        1
    \end{pmatrix}, \nonumber \\
    \ket{1 - f}=\ket{f'}=\frac{1}{\sqrt{2}}\begin{pmatrix}
        -1 \\
        i
    \end{pmatrix}. 
\end{align}
Here, $\ket{f},\ket{f'} \in \mathbb{C}^2 $.
These two above orthonormal eigenvectors provide the basis for the Hamiltonian and its complex conjugate. 
The eigenvalue equation Eq.~\eqref{feigenVeq} by virtue of the anticommutator constraint illustrates how the $H$ and $H^{\dagger}$ act as ladder operators between the Hermitian, orthonormal basis vectors $\ket{f}$ and $\ket{f'}$:
\begin{align}
    H\ket{f}&=e^{i\pi}\sqrt{f}\ket{f'}, \nonumber \\
    H\ket{f'}&=e^{-i\pi}\sqrt{1-f}\ket{f}, \nonumber \\
    H^{\dagger}\ket{f}&=e^{i\pi}\sqrt{1-f}\ket{f'}, \\
    H^{\dagger}\ket{f'}&=e^{-i\pi}\sqrt{f}\ket{f}. \nonumber
\end{align}
Following the above ladder operations, the Hamiltonian $H$ can be written in terms of the orthogonal basis vectors as
\begin{align}\label{HinHerbasis}
    H=\sqrt{f}e^{i\pi}\ket{f'}\bra{f}+\sqrt{1-f}e^{-i\pi}\ket{f}\bra{f'}.
\end{align}
In Eq.~\eqref{HinHerbasis}, it is shown that $H$ and consequently $H^{\dagger}$ take on an off-block diagonal structure in the Hermitian orthonormal basis $\ket{f},\ket{f'}$. In~\cite{bose2025PT}, we have discussed the $\eta$-pseudo Hermitian basis ($\ket{E_i^R}$,~$\ket{E_i^L}$), which is a basis folded by an $\eta$ metric such that the adjoint operation requires an $\eta$ operation in addition to complex conjugation. In that $\eta$-pseudo inner product space, the Hamiltonian takes a block diagonal form. 

The characteristic eigenvalue equation of the operator $M=H-E_{\pm}\mathbb{I}$ 
\begin{align}
    \text{det}(M)=\text{det}(H-E_{\pm}\mathbb{I})=0,
\end{align}
is solved to obtain the energy eigenvalues. 
We use the definition of the determinant of $M$,
\begin{align}
    \text{det}(M)&=\sum_{l,m}\epsilon_{lm} M_{0l}M_{1m}=0.
\end{align}
The Levi-Civita tensor picks permutations, i.e., $\epsilon_{01}=1, \epsilon_{10}=-1$ while $\epsilon_{00}=\epsilon_{11}=0$.
This results in
\begin{align}
    \text{det}(M)&=M_{00}M_{11}-M_{01}M_{10}=0 .
\end{align}
By calculating the matrix element of $M=H-E_{\pm}\mathbb{I}$ using~\eqref{HinHerbasis} and the completeness relation of the Hermitian basis $\mathbb{I}=\ket{f}\bra{f'}+\ket{f'}\bra{f}$,
we obtain the complex energy
\begin{align}\label{EeigenVaf}
    E_{\pm}=(f(1-f))^{\frac{1}{4}}\ge0,
\end{align}
where the magnitude of the energy is $f$ dependent. This also shows that $0\le f\le 1$. 
For the Hamiltonian in Eq.~\eqref{Hpt}, this yields the energy eigenvalue
\begin{align}\label{energyeigenVal}
    E_{\pm}=\pm\sqrt{\kappa(s^2-r^2)}.
\end{align}
It is essential that in Eq.~\eqref{Hpt}, $s\ge r$ to maintain real eigenvalues. This regime of values is named the $\pt$symmetric phase, as opposed to the broken symmetry phase, which contains imaginary eigenvalues.
The above eigenvalues guide us to obtaining the eigenvectors of $H$ in Eq.~\eqref{HinHerbasis}. Eigenvectors of $H^{\dagger}$ provide the left eigenvectors (dual space vectors) of the non-Hermitian Hamiltonian $H$. The relative coefficient $a=|a|e^{i\phi}$ of the two  basis vectors $\ket{f},\ket{f'}$ that make up the right and left eigenvectors are derived using the eigenvalue equation with corresponding energy eigenvalues, i.e., $H\ket{E_i^R}=E_i\ket{E_i^R}$ and $H^{\dagger}\ket{E_i^L}=E_i\ket{E_i^L}$, where $i\in\{+,-\}$. Assuming the global phase factor as one, we obtain
\begin{align}\label{Eneigenvec}
    \ket{E_{\pm}^R}&=\frac{1}{\sqrt{2}}(\ket{f}\pm|a|e^{\pm i\pi}\ket{f'}) \nonumber \\
    &=\frac{1}{\sqrt{2}}(\ket{f}\mp|a|\ket{f'}), \nonumber \\
    \ket{E_{\pm}^L}&=\frac{1}{\sqrt{2}}(\ket{f}\pm \frac{1}{|a|} e^{\pm i\pi} \ket{f'}) \nonumber \\
    &=\frac{1}{\sqrt{2}}(\ket{f}\mp \frac{1}{|a|}\ket{f'}).
\end{align}
Here $a=|a|e^{i\phi}$, where $|a|=(\frac{f}{1-f})^{\frac{1}{4}}$. For the Hamiltonian in Eq.~(\ref{Hpt}), $|a|=\sqrt{\frac{s+r}{s-r}}$ and $\phi=\pi$. It is noteworthy that the choice of $\ket{f'}$ in Eq.~(\ref{feigenve}) is not unique. A prime choice is 
\begin{align}
    \ket{f'}=\frac{1}{\sqrt{2}}\begin{pmatrix}
        1 \\
        -i
    \end{pmatrix} .
\end{align}
This choice is characterized by the azimuthal angle $\phi=0$. One has $0\le\ \phi \le2\pi$ degrees of freedom  to choose  $\ket{f'}$ while maintaining orthogonality with $\ket{f}$.
From Eq.~\eqref{Eneigenvec} it is evident that
\begin{align}\label{orthonormality}
    \bra{E_i^L}E_{j}^R\rangle=\delta_{ij},
\end{align}
where $i,j=\pm$. This establishes the orthogonality of the basis vectors $\ket{E_i^R}, \ket{E_j^L}$ of the Hamiltonian $H$. 
\subsection{The broken phase of complex eigenvalues}
The dynamics of the broken phase of the $\pt$symmetric Hamiltonian is analogous to a particle trapped in a finite potential well, where the norm of a state undergoing evolution of the $\pt$symmetric evolution $\bra{\psi(t)^L}\psi(t)^R\rangle$ decays exponentially owing to the imaginary part of the complex eigenvalues. The norm $\bra{\psi(t)^L}\psi(t)^R\rangle$ denotes the probability that the particle has not escaped the potential well by quantum tunneling. For details, see~\cite{Brody_2014,Brody2016}.
When the eigenvalues of the $\pt$symmetric Hamiltonian are not real, i.e., when the system is in the broken phase, then the $\eta$-Hermiticity ~\cite{bose2025PT} does not hold, i.e.,
        \begin{align}
            H^{\ddagger}=\eta^{-1}H^{\dagger}\eta=\sum_nE_n^* \ket{E^R_n}\bra{E^L_n}\neq H.
        \end{align}
In section IV of~\cite{kleefeld2009constructiongeneralinnerproduct} and in~\cite{kleefeld2023equivalencepaisuhlenbeckoscillatormodel}, the problem of complex eigenvalues is dealt with in a different way. For two non-Hermitian harmonic oscillators, the concept of `PT completion' is introduced by adding to the Hamiltonian the complex conjugate of the same, making it a sum of a holomorphic and anti-holomorphic term in different Hilbert subspaces (causal and anti-causal Hamiltonians). As a consequence the eigenvalues for the right eigenstates $\ket{(m,n)^R}$ or left eigenstates $\bra{(m,n)^L}$ of the total  Hamiltonian becomes $E_{m,n} = E_m + (E_n)^\ast$, i.e., the sum of the complex valued eigenvalues of the holomorphic Hamiltonian and the complex conjugate eigenvalues of the antiholomorphic Hamilonian. For the real eigenvalue case, the biorthonormal vectors $\ket{(m,n)^R}$ reduce to $\ket{n^R}$, denoting a special case $m=n$. However, this approach is technically very different from what we intend to do, and we keep the discussion of the broken $\pt$symmetry phase for our future endeavors, hoping that it would lead to a fully non-Hermitian generalization of quantum theory.

\subsection{Exceptional point and Normal point}
The eigenvalue $f$ of the Hermitian $F$ [Eq.~(\ref{feigenVeq})] satisfying $0\le f \le 1$ has different regimes of values where the Hamiltonian $H$ exhibits salient features. At the endpoints, we have the vacua $f=0,1$, and at the middle point we have $f=1/2$.
For $f=0,1$, the magnitude of the energy eigenvalue in Eq.~\eqref{EeigenVaf} is the same and is equal to 0, and the relative coefficient between the two basis vectors that make up the eigenvectors is $|a|=0$ (when $f=0$) and $|a|\rightarrow\infty$ (when $f=1$). For $|a| = 0$, the energy eigenvectors $\ket{E_\pm^R}$ collapse to $\ket{f}$ and the dual eigenvectors blow up, violating the orthonormality of the eigenvectors in Eq.~\eqref{orthonormality}. A similar case happens when $|a| \to \infty$. Therefore, this behavior could be called the exceptional point. 
For the $\pt$Symmetric Hamiltonian in Eq.~\eqref{Hpt}, the exceptional point is $f=0$ when $r=-s$ and is $f=1$ when $r=s$. Thus, at these exceptional points, the Hamiltonians boil down to
\begin{align}\label{exchermi}
    H= \frac{1}{2}\begin{pmatrix}
    \pm i & 1 \\
    1 & \mp i
    \end{pmatrix}.
\end{align}
At the middle point of the parameter space $0\le f \le 1$ we have $r=0$ (making the imaginary elements vanish) and $~s\geq r=0$ (condition of the unbroken phase for real eigenvalues) for $f=1/2$. At this point, the energy eigenvalue [Eq.~\eqref{EeigenVaf}] is a positive number. 
Since $|a|=\sqrt{\frac{s+r}{s-r}}=1$ at this normal point, the eigenvectors now contain the two Hermitian basis vectors in equal proportions.
Interestingly, at the normal point, the Hamiltonian is Hermitian
\begin{align}\label{hermi}
    H= \frac{1}{\sqrt{2}}\begin{pmatrix}
    0 & 1 \\
    1 & 0
    \end{pmatrix},
\end{align}
with equal left and right eigenvectors.
The distance of the $f$ values from the point $f=1/2$ relates to the degree of non-Hermiticity of the Hamiltonian, tunable in terms of the eigenvalue of the Hermitian operator $F$.

\section{Ergotropy for a $\pt$Symmetric Hamiltonian}\label{SecIII}
In quantum thermodynamics, ergotropy quantifies the maximum amount of work that can be extracted from a state $\rho(t)$  at time $t$,
by unitary operations~\cite{Allahverdyan_2004}. For a non-Hermitian system, more precisely an $\eta$-pseudo Hermitian $\pt$Symmetric Hamiltonian, we redefine the density matrix of the system in the $\eta$-pseudo inner product space as the generalized density matrix $\rho_G(t)$. In other words, in this bi-orthonormal basis, the projectors that make up the matrix elements of $\rho_G(t)$ are now constructed by the corresponding right ($\ket{E_i^R}$) and left(dual)($\ket{E_i^L}$) eigenvectors. For the Hamiltonian we work with, i.e., $H$, these are provided in Eq.~\eqref{Eneigenvec}.
The ergotropy for the $\pt$symmetric $H$ is
\begin{align}\label{ergotropy}
    \mathcal{W}=\text{Tr}(H\rho_G(t))-\text{Tr}(H\rho^\mathrm{p}),
\end{align}
where $\rho^{\mathrm{p}}$ is the passive state associated with $\rho_G(t)$.
A comprehensive derivation of ergotropy suitable for bi-orthonormal eigenspace and $\eta$-unitary evolution is presented in Appendix E.

A passive state is obtained by rearranging the eigenvalues of $\rho_G(t)$ in decreasing order and aligning them with the increasing energy eigenvalues of 
$H$. By construction, no work can be extracted from 
$\rho^{\mathrm{p}}$ under $\eta$-pseudo unitary evolution.
\subsection{Calculation of $\rho_G(t)$ in the Hermitian basis}
The dual space that complements the eigenspace of $H$ is not obtained by mere complex conjugation. In our previous work~\cite{bose2025PT}, we have shown that an additional $\eta$-metric is necessary to map the eigenspace to its dual~\cite{kleefeld_2009arXiv0906.1011K, SimilarityinPT,similarity2Mostafazadeh2003,similarity3, SimilarityAshida02072020, SimilarityinPT, Scolarici2006}. In this work, we take a different approach to expressing both $H$ and $H^{\dagger}$  in a Hermitian basis ($\ket{f},\ket{f'}$) that spans them both. Using this approach, analytical calculations of the thermodynamic quantities become easier. Equation \eqref{Eneigenvec} shows how the bi-orthonormal eigenvectors are expressed in terms of the Hermitian basis vectors.  In this space, the projection operators that constitute the density matrix are modified such that
\begin{align}\label{rhoGt}
    \rho_G(t)=\sum_{ij}\rho^{ij}(t)\ket{E_i^R}\bra{E_j^L}=\sum_{ij}\rho^{ij}(t)\ket{E_i^R}\bra{E_j^R}\eta.
\end{align}
The elements $\rho^{ij}(t)$'s are the coefficients of the projectors, i.e., the information of the state, including population and coherence terms. The projection operators are defined as,
\begin{align}
    \Pi_{ij}=\ket{E_i^R}\bra{E_j^L}.
\end{align}
These projection operators project the $i,j$-th element of any operator from the biorthonormal space. The origin of the left eigenvectors of the $\pt$symmetric Hamiltonian is described in Appendix B.
The role of the Hermitian $\eta $ metric as a map to the dual space is clear from the above equation. More explicitly, $\eta\ket{E^R}=\ket{E^L}$, so that the inner product yields
\begin{align}\label{etapseudoinnerproduct}
    \bra{E_i^R}\eta\ket{E_j^R}=\bra{E_i^L}E_j^R\rangle=\langle E_i^R\ket{E_j^L}=\delta_{ij},
\end{align}
where
\begin{align}\label{eta_definition}
    &\eta=\sum_n \ket{E_n^L}\bra{E_n^L},
\end{align}
attests to its Hermiticity~\cite{bose2025PT}. 
The calculation of $\rho_G(t)$ is elaborated in Appendix A.

\subsection{Modified measurement postulates}
The quantum mechanical observables of a Hermitian evolution are generalized to this $\pt$symmetric case by introducing the metric $\eta$ and redefining the inner product~\eqref{etapseudoinnerproduct}.
For the two-dimensional Hilbert space of the $\pt$symmetric Hamiltonian, we have discrete eigenvalues of energy along with two left and their corresponding right eigenvectors connected by $\ket{E_i^L}=\eta \ket{E_i^R}$, $\eta$ is specified by the parameters of the Hamiltonian. To express an operator, the projectors of the operator must be modified. This is seen clearly in the spectral decomposition,
\begin{align}
     H=\sum_{n}E_n \ket{E_n^R}\bra{E_n^L}.
\end{align}
The spectral decomposition is the consequence of the modified completeness relation,
\begin{align}\label{eq_complete}
    &\sum_n\ket{E_n^R}\bra{E_n^L}=\mathbb{I}.
\end{align}
Any quantum mechanical observable suitable for a $\pt$symmetric evolution can be written as a projective measurement operator involving the modified projectors. These observables have a discrete spectrum depending on the Hilbert space dimension.
We can also define a projective measurements in this $\eta$-inner product space simply as the projectors itself,
\begin{align}
    M_i=\ket{E_i^R}\bra{E_i^L}.
\end{align}
This also obeys,
\begin{align}
    M_iM_j&=\ket{E_i^R}\bra{E_i^L}{E_j^R}\rangle\bra{E_j^L}=\delta_{ij}M.
\end{align}
The $M_i$s are $\eta$-Hermitian, i.e.,
\begin{align}
    M_i^{\ddagger}&=\eta^{-1}M_i^{\dagger}\eta, \nonumber \\
    &=\eta^{-1}(\ket{E_i^R}\bra{E_i^L})^\dagger\eta, \nonumber \\
    &=\eta^{-1}(\ket{E_i^L}\bra{E_i^R})\eta, \nonumber \\
    &=\ket{E_i^R}\bra{E_i^L}=M_i
\end{align}
As elaborated in~\cite{bose2025PT}, the trace operation in this $\eta$-inner product space corresponds to the action $\sum_i\bra{E_i^L}.\ket{E_i^R}$. This preserves the unity of trace of the generalized density matrix, $\sum_k \bra{E_k^L} (\rho_G(t))\ket{E_k^R}=\sum_k \bra{E_k^L} (\sum_{ij} \rho_{ij}(t) \ket{E_i^R}\bra{E_j^L})\ket{E_k^R}=\sum_k \rho_{kk}(t)=1$, where $\rho_{kk}$s are the information theoretic population terms.
To illustrate the modified measurement postulates, we take a normalized quantum state evolving under the $\eta$-Hermitian $\pt$symmetric Hamiltonian as $\ket{\psi^R}=\sum_n c_n\ket{E_n^R}$ and its $\eta$ adjoint dual state $\bra{\psi^L}=\sum_n c^*\bra{E_n^L}$ rendering $\bra{\psi^L}\psi^R\rangle=1$ following the $\eta$-inner product. The probability of getting a certain result $n$ is given by,
\begin{align}
    p(n)=\bra{\psi^L}M_i\ket{\psi^R}=|c_n|^2,
\end{align}
and the state after measurement of operator $M_i$ is given by,
\begin{align}
    \frac{M_i\ket{\psi^R}}{\sqrt{\bra{\psi^L} M_i \ket{\psi^R}}}.
\end{align}
After measurement, the state is projected to one of the modified eigenvectors.
The invariance of average energy in this formulation is elaborated in Appendix F.

\subsection{Calculation of ergotropy}
Once $\rho_G(t)$ is obtained from Eq.~\eqref{rhoGt}, its eigenvalues are calculated and compared. The larger (smaller) population is put in the lower (higher) energy eigenstates, such that,
\begin{align}
    \rho^{\mathrm{p}}=\lambda_+\ket{E_+^R}\bra{E_+^L}+\lambda_-\ket{E_-^R}\bra{E_-^L}.
\end{align}
Since $E_+>E_-$, then $\lambda_+<\lambda_-$ should be the choice for the construction of a passive state.
We can now calculate the ergotropy following Eq.~\eqref{ergotropy} for a closed system with a $\pt$Symmetric Hamiltonian Eq.~\eqref{Hpt},
\begin{align}
    \mathcal{W}&=|E|(\rho^{22}(t)-\rho^{11}(t))-|E|(\lambda_--\lambda_+) \nonumber \\
    &=|E|(\rho^{22}(t)-\rho^{11}(t)-\lambda_-+\lambda_+) \nonumber \\
    &=|E|(1-2\rho^{11}(t)-\lambda_-+\lambda_+),
\end{align}
where $|E|=\sqrt{\kappa(s^2-r^2)}$ is the absolute value of the eigenvalue, Eq.~(\ref{energyeigenVal}). The eigenvalues of $\rho_G(t)$ are,
\begin{widetext}
\begin{align}\label{RhoGeigenvalue}
    \lambda_\pm=\frac{1}{2}\pm\frac{1}{2}\sqrt{1-\frac{4}{s^2-r^2} \left\{ rs\rho^{21}(t)+s^2(\rho^{11}(t)\rho^{22}(t)-\rho^{12}(t)\rho^{21}(t))-r^2\rho^{11}(t)\rho^{22}(t)\right\}}.
\end{align}
\end{widetext}
The eigenvalues naturally depend on the geometry of the Hilbert space, denoted by the parameters of the Hamiltonian $r,~s$, and also on the population and coherence terms of the evolved quantum state. 

\section{Ergotropy of a $\pt$symmetric Hamiltonian as an open system}\label{SecIV}
\begin{figure}
    \centering
    \includegraphics[width=1\linewidth]{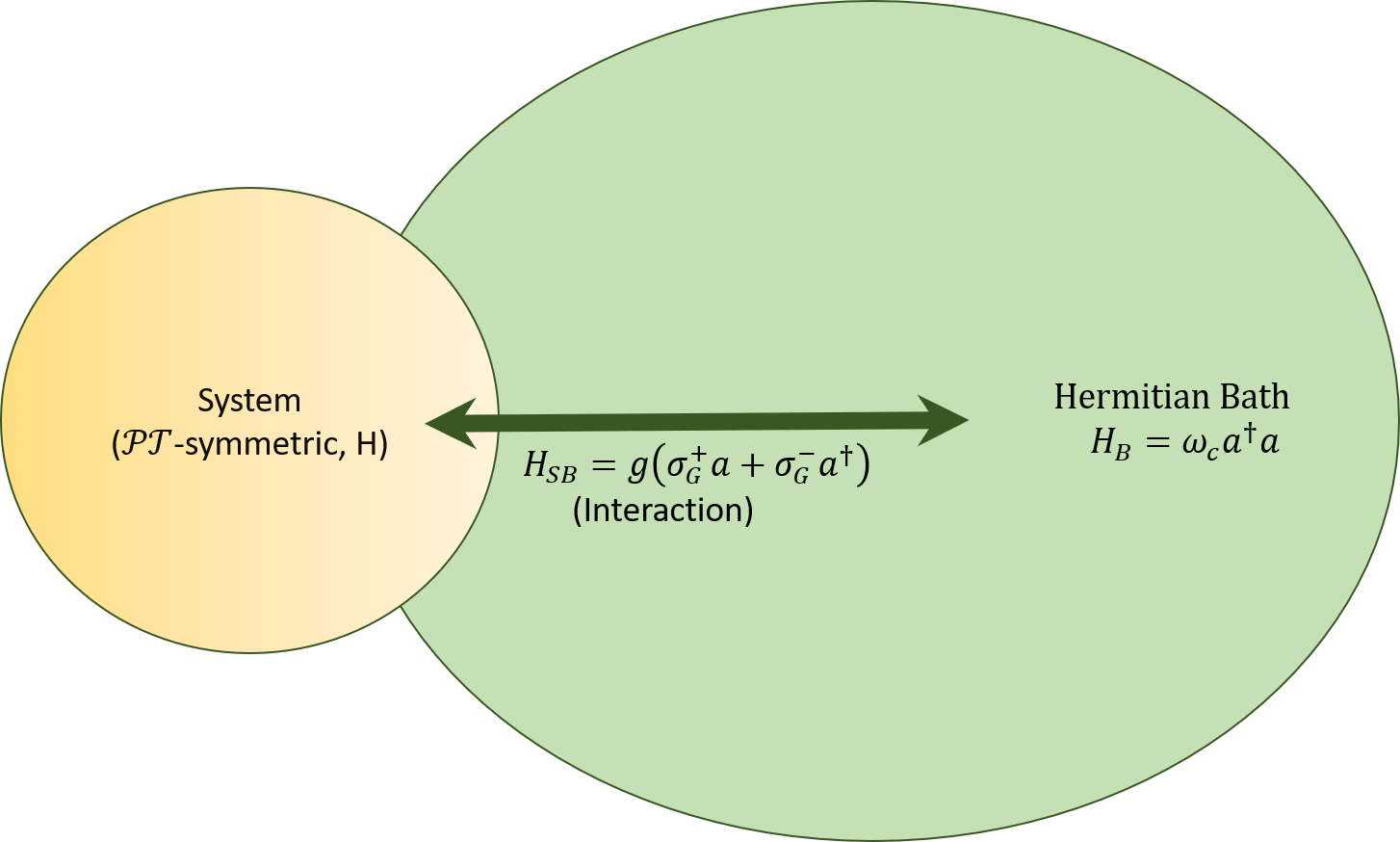}
    \caption{A diagrammatic representation of the open system scheme we apply for a $\pt$symmetric Hamiltonian.}
    \label{fig:placeholder}
\end{figure}
To understand the ergotropy in the presence of the open system effects, the system is connected to a simple bath. We consider a scenario where a single-mode bosonic bath of frequency $\omega_c$ is coupled to the $\mathcal{PT-}$symmetric Hamiltonian $H$. Inspired by light-matter interaction models, e.g., the Jaynes-Cummings model~\cite{jaynes_cummings}, the corresponding composite system-bath Hamiltonian can be envisaged to be
\begin{align}\label{HUtilde}
    \tilde{H}_U&=H+H_B+g(\sigma^+_G\otimes a + \sigma^-_G \otimes a^{\dagger}),
\end{align}
where
\begin{align}\label{eq_bath_ham}
    H_B&=\omega_ca^{\dagger}a.
\end{align}
Here, $H_B$ is the Hermitian Hamiltonian of the bath and $g$ is the coupling constant taken to be real.
In the $\eta$-pseudo inner product space, Eq.~\eqref{etapseudoinnerproduct}, the adjoint of an operator is acquired by rotating the complex conjugate of the given operator by the Hermitian metric $\eta$. The $\pt$symmetry of the Hamiltonian in Eq.~\eqref{Hpt} is also manifested in its $\eta$-pseudo Hermiticity,
\begin{align}
    H^{\ddagger}=\eta^{-1}H^{\dagger}\eta=H,
\end{align}
where the double dagger relates to the $\eta$ adjoint of $H$.
The knowledge of $\eta$ is gained with the knowledge of left and right eigenvectors, Eq.~(\ref{eta_definition})~\cite{bose2025PT}.
We calculate the reduced generalized density matrix for an open system as
\begin{align}\label{rhoGt_eta_uni}
    \rho_G(t)=\text{Tr}_B[U\left\{\rho_G(0)\otimes \rho_B(0)\right\}U^{\ddagger}].
\end{align}
Appendix C contains the details of the $\eta$ adjoint $U^{\ddagger}$ of the $\eta$unitary operator $U$.
In general, the dynamics of $\rho_G(t)$ is non-Markovian in nature~\cite{bose2025PT}. 
$\tilde{H}_U$, Eq.~\eqref{HUtilde}, is $\eta$-pseudo Hermitian, i.e.,
\begin{align}\label{etahermiofComposite}
    (\eta^{-1}\otimes I_B)\tilde{H}_U^{\dagger}(\eta \otimes I_B)=\tilde{H}_U^\ddagger = \tilde H_U,
\end{align}
following the $\eta$-pseudo Hermiticity of $H$, i.e., $\eta^{-1}H^{\dagger}\eta=H$ and the fact that $\sigma^+_G$ and $\sigma^-_G$, the system operators suitable for the $\eta$-pseudo inner product space,
\begin{align}\label{Biortho_sigmas}
    \sigma^+_G &=\ket{E_+^R}\bra{E_-^L}, \nonumber \\
    \sigma^-_G &=\ket{E_-^R}\bra{E_+^L} ,
\end{align}
are $\eta$-pseudo adjoints of each other, as shown below
\begin{align}
    (\sigma^+_G)^{\ddagger}&=\eta^{-1}(\sigma^+_G)^{\dagger}\eta =\eta^{-1}\ket{E_+^L}\bra{E_-^R}\eta \nonumber \\
    &=\eta^{-1}\eta\ket{E_+^R}\bra{E_-^L} = \ket{E_+^R}\bra{E_-^L} =\sigma^-_G.
\end{align}
Redefinitions of Pauli matrices in the bi-orthonormal basis are illustrated in Appendix D.
Three different initial system states are taken: the $\mathcal{PT-}$symmetric ground state ($\rho_{G}^1$), excited state ($\rho_{G}^2$), and an intermediate state ($\rho_{G}^3$),
\begin{align}\label{Initialstates}
    \rho_{G}^1&=\ket{E_-^R}\bra{E_-^L}, \nonumber \\
    \rho_{G}^2&=\ket{E_+^R}\bra{E_+^L}, \nonumber \\
    \rho_{G}^3&=\frac{3}{4}\ket{E_-^R}\bra{E_-^L}+\frac{\sqrt{3}}{4}\ket{E_+^R}\bra{E_-^L}\nonumber \\
    &+\frac{\sqrt{3}}{4}\ket{E_-^R}\bra{E_+^L}+\frac{1}{4}\ket{E_+^R}\bra{E_+^L}.
\end{align}
The initial state of the bath is taken as the thermal state $\rho_B(0)=e^{-H_B/T}/\text{Tr}\left( e^{-H_B/T} \right)=e^{-\beta H_B}/\text{Tr}\left(e^{-\beta H_B}\right)$.
The $\eta$-pseudo adjoint of the $\eta$-unitary operator ~\cite{Scolarici2006,Scolarici2007,bose2025PT} is
\begin{align}\label{Uddager}
    U^{\ddagger}&=e^{i\tilde{H}_U t}.
\end{align}
Given the $\pt$symmetric initial states and a $\eta$-pseudo Hermitian Hamiltonian of the total evolution, we numerically calculate ergotropy ($\mathcal{W}$) following Eq.~\eqref{ergotropy}. The passive state $\rho^{\mathrm{p}}$ is also calculated numerically in this case.
\begin{figure}
    \centering
    \includegraphics[width=1\linewidth]{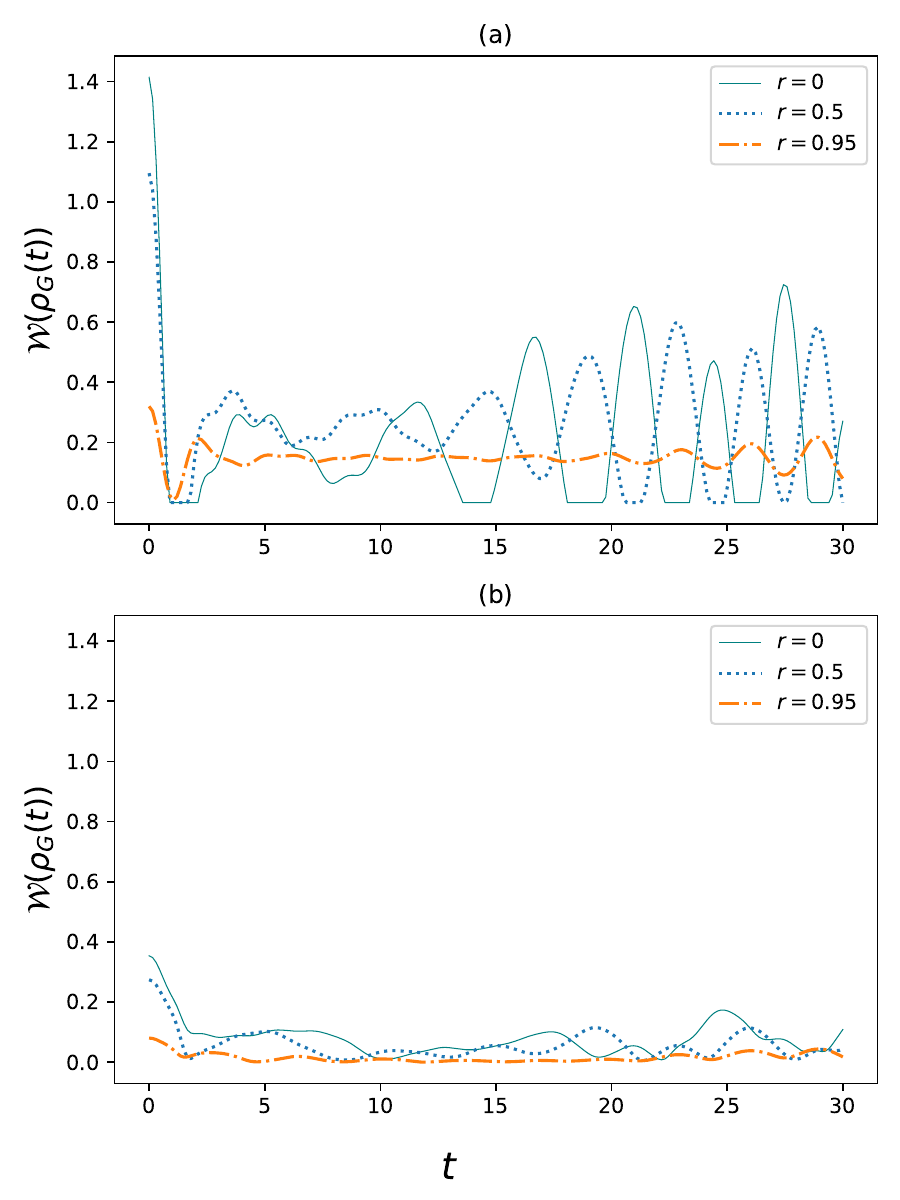}
    \caption{Variation of ergotropy $\mathcal{W}(\rho_G(t))$ in an open system. The initial states are the excited state $\rho_G^2$ in (a), the state $\rho_G^3$ in (b), defined in Eq.~\eqref{Initialstates}. The Hamiltonian parameter $s$ is kept fixed at one, and $r$ is varied to render different regimes of Hermiticity. The coupling constant is $g=0.5$, and the bath frequency, dimension, and temperature are consecutively $\omega_c=2,~d_B=15,~T=10$. }
    \label{fig:Ergotropy_PTSymmetry}
\end{figure}
In Fig.~\ref{fig:Ergotropy_PTSymmetry}, we see that the ergotropy starts at a maximum value when the initial state is the excited state. In the intermediate state, the ratio of the population is more toward the ground state, and this gives a comparatively low initial value of $\mathcal{W}$. When the initial state is the ground state (not shown in the figure), the ergotropy is zero in comparison to the rest, indicating the fact that when one starts with the system in the ground state, there is no effective extractable work. 
Coupling to the radiation bath causes the ergotropy of the $\pt$symmetric system to decrease in a featureless oscillatory manner. 
It is noteworthy that, except for the case when the initial state is the $\pt$symmetric ground state, the initial or the closed system ergotropy is maximum when the $\pt$symmetry is relaxed, and the $H$ in Eq.~\eqref{Hpt} is in its Hermitian limit at the normal point in the parameter space of the Hamiltonian and reduces to Eq.~\eqref{hermi}. Near the exceptional point, $r\simeq s$, the initial ergotropy is the lowest for all the cases when it approaches the `vacua' or the end of the parameter space. From Eq.~\eqref{Eneigenvec} it is clear that, at $f=1,0$  when the relative coefficient $|a|=\left(\frac{f}{1-f}\right)^{\frac{1}{4}}$ is $\infty,0$, the orthonormality of the eigenvectors no longer holds, i.e., $\bra{E_i^L}E_i^R\rangle\neq1$. This can be the reason for diverging values of ergotropy in the exceptional points of the Hamiltonian. In Fig.~\ref{fig:Ergotropy_PTSymmetry} plots, the dot-dashed curves show the case of $s=1,r=0.95\simeq s $ and not $r=s$ to show how ergotropy behaves when it approaches the exceptional point before diverging.
In all the plots, we see late-time revivals of ergotropy ($\mathcal{W}$). Since we work with an open system involving a single-mode bosonic bath, these revivals can be attributed to the non-Markovian effects of the open system's evolution.
Studying the plots, it can be inferred that, as long as ergotropy is concerned, it decreases as the Hamiltonian becomes more non-Hermitian, which in our case is the $\pt$symmetry. At its Hermitian limit, the Hamiltonian reduces to a Pauli $\sigma_x$ type system, and in this case, the ergotropy is higher than the case when it is at its non-Hermitian regime, as can be observed from the $r=0$ solid line in Fig.~\ref{fig:Ergotropy_PTSymmetry}. Relaxation from $\pt$symmetry to Hermiticity causes a gain in ergotropy.
\section{Thermodynamics of a $\pt$symmetric quantum system}\label{SecV}
In the last section, we have modeled a quantum thermodynamic scenario where the system is represented by a $\pt$symmetric Hamiltonian, Eq.~\eqref{Hpt}, and the bath is represented by a Hermitian Hamiltonian $H_B$, Eq.~\eqref{eq_bath_ham}. In this section, we verify the three laws of thermodynamics for this model. The first law of thermodynamics is,
\begin{align}
    dU=dW-dQ_B,
\end{align}
where $dU$ is the change in the internal energy of the system, $dW$ is the work done in the process of evolution by $H$, and $dQ_B$ is the amount of heat exchanged between the system and the bath in that process. They are defined for an open system as,
\begin{align}
    dU&=\text{Tr}[H(\rho_{G}(t)-\rho_{G}(0))], \nonumber \\
    dW&=\text{Tr}[H_{GB}(\rho_{GB}(0)-\rho_{GB}(t))], \nonumber \\
    dQ_B&=\text{Tr}[H_B(\rho_{B}(t)-\rho_{B}(0))],
\end{align}
where $H_{GB} = g (\sigma_G^+a + \sigma^-_Ga^\dagger)$ is the interaction Hamiltonian. 
Here, $\rho_G(t)$ is the system density matrix for the evolution of the $\pt$ symmetric Hamiltonian, expressed on a bi-orthonormal basis. $\rho_{GB}$ stands for the composite density matrix of the bi-orthonormal system and the Hermitian bath. $\rho_B$ is the density matrix of the bath obtained by partial tracing the degrees of freedom of the system. The detail of how the work transfer takes the above form is described in Appendix G.
It is observed, Fig.~\ref{fig:FirstLaw_PTSymmetry}, that for all regimes of the Hamiltonian $H$, the first law is obeyed. 
\begin{figure}
    \centering
    \includegraphics[width=1\linewidth]{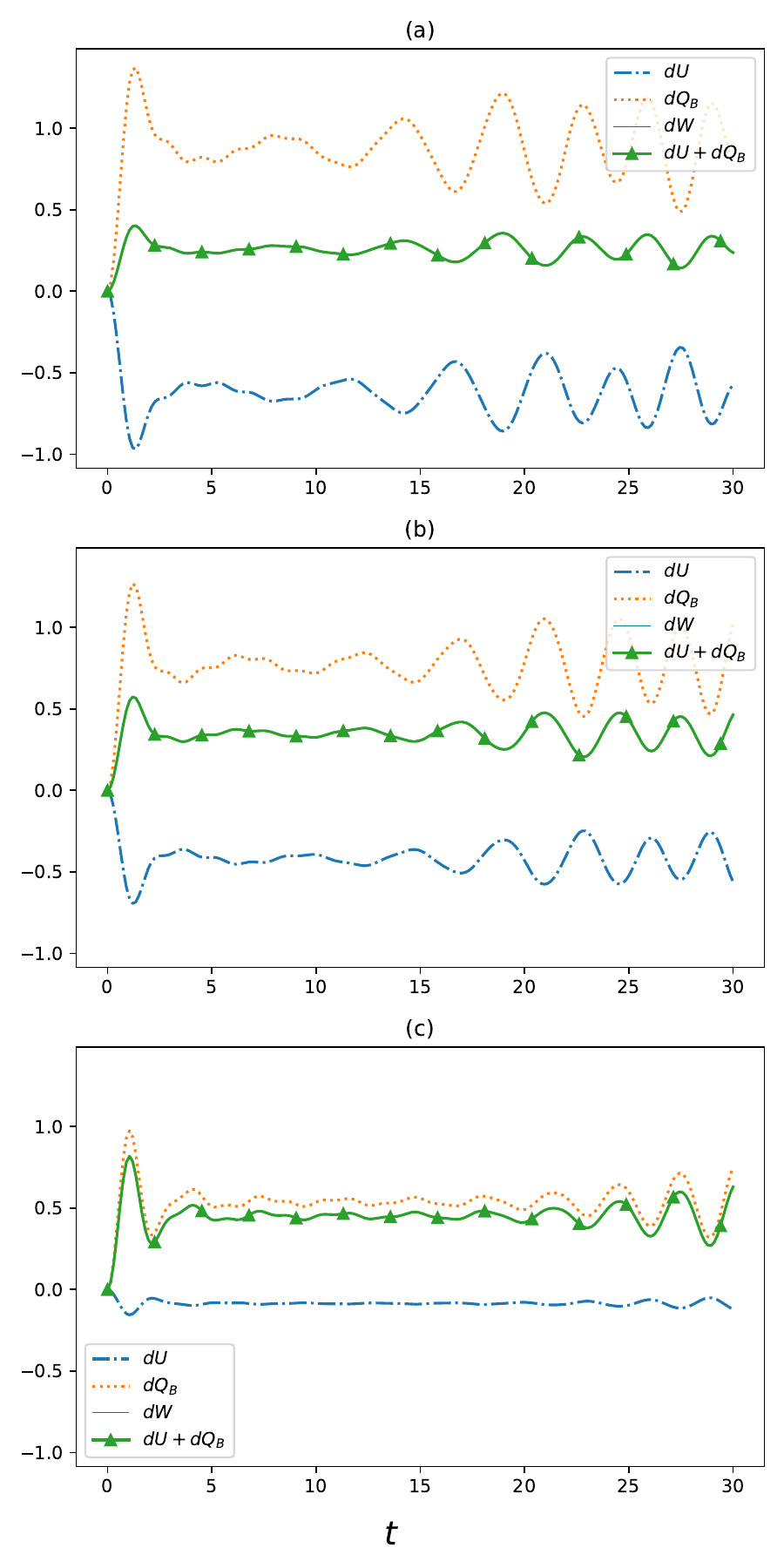}
    \caption{Different thermodynamic quantities are plotted when the initial state is the excited state $\rho_G^2$. (a) is for the Hermitian limit at $r=0$, (b) is for the non-Hermitian Hamiltonian at $r=0.5$, and (c) is for when the Hamiltonian approaches the exceptional limit, $r=0.95$. All cases have $s=1$. The coupling constant is $g=0.5$, and the bath frequency, dimension, and temperature are consecutively, $\omega_c=2,~d_B=15,~T=10$}
    \label{fig:FirstLaw_PTSymmetry}
\end{figure}
The second law of thermodynamics in the case of a quantum system ensures the positivity of the entropy production ($\Sigma$),
\begin{align}\label{ent_prod}
    \Sigma=\mathcal{S}[\rho_{GB}(t)||\rho_G(t)\otimes \rho_B(0)]\ge0.
\end{align}
Here $\mathcal{S}(\rho||\sigma)=\text{Tr}(\rho \log \rho -\rho \log \sigma )$ is the relative entropy between two density matrices $\rho$ and $\sigma$.
\begin{figure}
    \centering
    \includegraphics[width=1\linewidth]{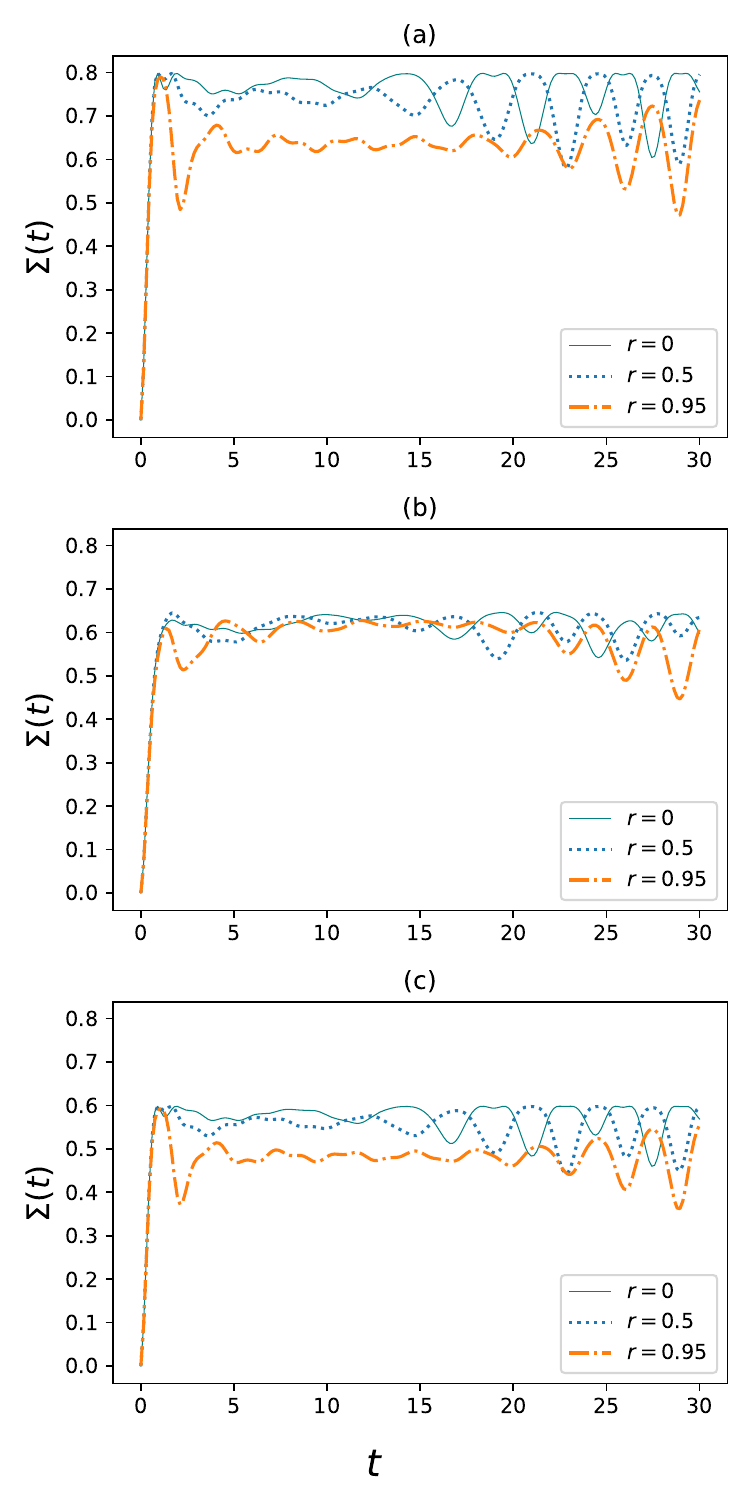}
    \caption{Entropy production ($\Sigma$) for the $\pt$symmetric open quantum system. The initial states [Eq.~\eqref{Initialstates}] are the excited state $\rho_G^2$ in (a), the state $\rho_G^3$ in (b), and the ground state $\rho_G^1$ in (c). The parameter $s$ is kept fixed at one, and $r$ is varied to render different regimes of non-Hermiticity. The coupling constant is $g=0.5$, and the bath frequency, dimension, and temperature are consecutively, $\omega_c=2,~d_B=15,~T=10$.}
    \label{fig:Entropy_prod_PTSymmetry}
\end{figure}

In Fig.~\ref{fig:Entropy_prod_PTSymmetry} we see that for all initial states and for all regimes the entropy production ($\Sigma$) is positive. After a sharp initial increase, the entropy production exhibits much smaller oscillatory behavior. Interestingly, $\Sigma$ shows a sharper dip after an initial rise when the initial state is the excited state or the ground state, which are coherence-less states. For the state $\rho_G^3$, this dip is not so sharp, and when the Hamiltonian approaches the exceptional point, the $\Sigma$ profile is qualitatively similar to the profiles at other regimes. For $\rho_G^1,~\rho_G^2$, the behavior of $\Sigma$ near the exceptional points is more distinct; more precisely, it runs lower than that of the other regimes.

\begin{figure}
    \centering
    \includegraphics[width=1\linewidth]{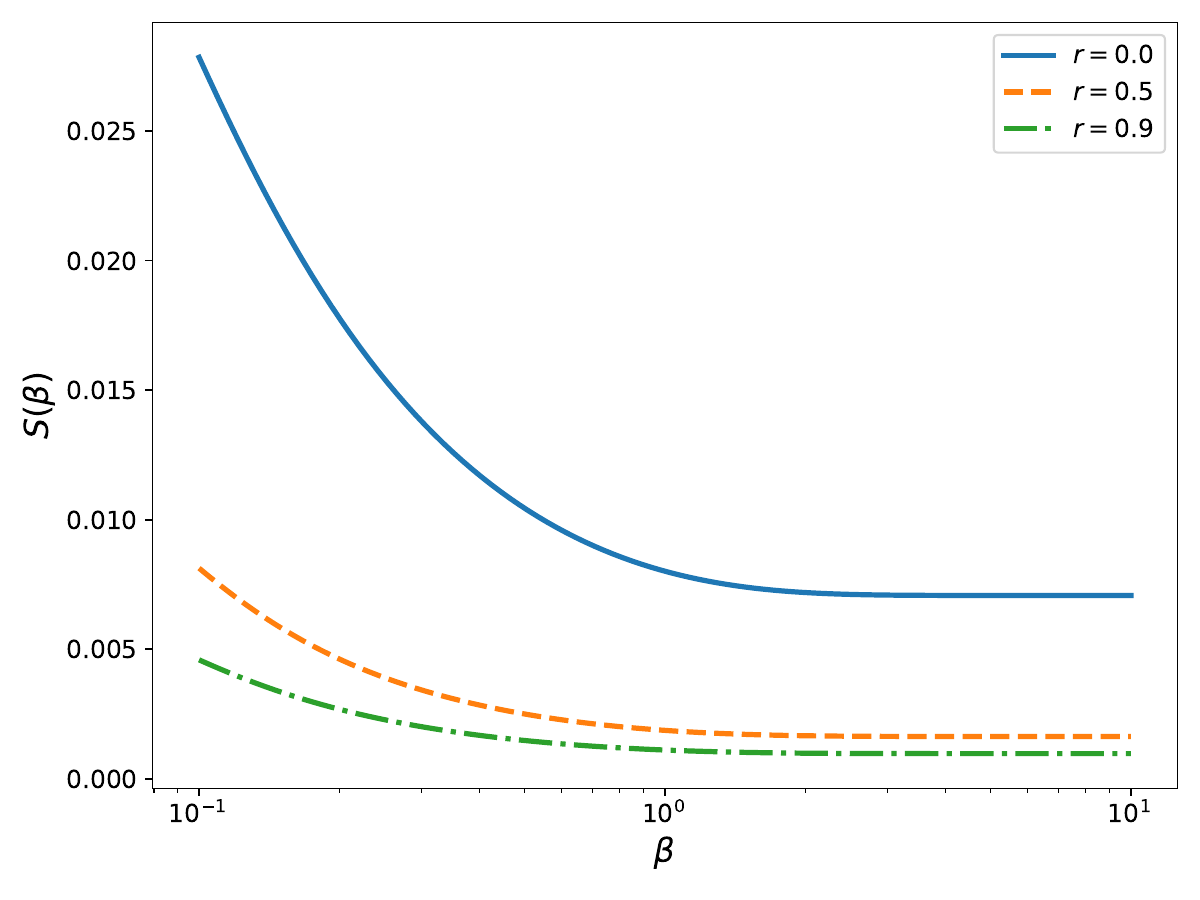}
    \caption{The von-Neumann entropy of the generalized density matrix at coupling constant $g=0.01$ in the long time limit, for different values of $r$, while $s=1$ with the variation of the inverse temperature $\beta$ of the bath.}
    \label{fig:Entropy_limit_PTSymmetry}
\end{figure}

In the third law,
\begin{align}
    S(T)|_{T\to 0}=-\text{Tr}(\rho_G(t)\log\rho_G(t))|_{T\to 0} \to 0,
\end{align}
in the long time limit, where $S(T)$ is the von-Neumann entropy. For all initial states and in different parameter ranges of the Hamiltonian $H$, the above relation is obeyed when the coupling constant with the bath $g$ is considerably low. For stronger couplings, the von-Neumann entropy gets modified by a temperature-dependent effective term~\cite{Miller2018_inbook}. 
Fig.~\ref{fig:Entropy_limit_PTSymmetry} shows the analysis of $S(T)$ at very low temperature, and it can be observed that the von-Neumann entropy approaches zero as the temperature decreases.

Furthermore, we want to emphasize here that though the thermodynamic laws developed in this article are demonstrated for a single-mode harmonic oscillator bath, the results are general and are applicable to a regular, large-sized bath. Here, the single-mode bath has been chosen for numerical ease in computing observables that require information about the bath state, such as the system's entropy production. Nevertheless, we can show that these quantities scale well even if we increase the bath size. To demonstrate this, we calculate the entropy production of the $\pt$symmetric system by attaching a finite bath to the system~\cite{Fluctu1, tiwari2024strong}.

\subsection{Entropy production of the $\pt$symmetric system interacting with a central spin bath}
\begin{figure}
    \centering
    \includegraphics[width=1\linewidth]{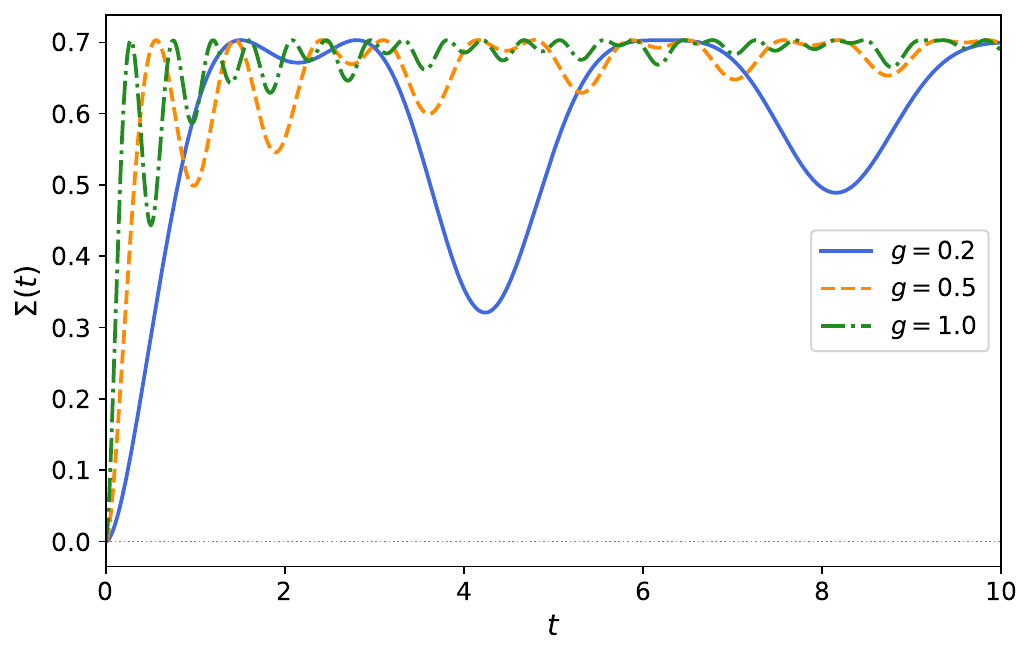}
    \caption{Variation of the entropy production for the two-level $\pt$symmetric system interacting with the finite-sized central spin bath, Eq.~\eqref{pt_cs}. The parameters are $r = 0.1, s = 0.4, \omega = 1.0, T = 1, N = 50$.}
    \label{fig_entropy_production_CS_model}
\end{figure}

Here, we attach a central spin bath to the two-level $\pt$symmetric system~\cite{tiwari2024strong}. The bath is composed of $N$ identical two-level systems interacting uniformly with the $\pt$symmetric system. The total Hamiltonian of the system is given by
\begin{align}\label{pt_cs}
    \tilde H_{\rm CS} = H + \frac{\omega}N J_z + \frac{g_{cs}}{\sqrt{N}}\left(\sigma_G^+J_- + \sigma^-_GJ_+\right), 
\end{align}
where $\omega$ is the frequency of the bath spins and $g_{cs}$ is the coupling constant for the interaction between the system and the bath. $H$ is the two-level $\pt$symmetric system, Eq.~\eqref{Hpt}. The evolution of the system is governed by a similar $\eta-$unitary evolution as discussed around Eq.~\eqref{rhoGt_eta_uni}. The initial state of the system and the bath are taken to be $\rho_G^2$ [Eq.~\eqref{Initialstates}], and $e^{-\beta\omega J_z/N}/{\rm Tr}\left[e^{-\beta\omega J_z/N}\right]$, respectively. The entropy production is calculated using Eq.~\eqref{ent_prod} and is plotted in Fig.~\ref{fig_entropy_production_CS_model}.
The figure shows that the entropy production is positive throughout for this model, verifying the second law of quantum thermodynamics. The rise in the entropy production is steeper for higher values of $g$. The bath consists of 50 spins, indicating that the formalism holds as we scale the bath size.

\section{Conclusions}\label{SecVI}
In this paper, we have selected a subclass of $\pt$symmetric Hamiltonians obeying an anti-commutator relation with their complex conjugate to devise a Hermitian basis that spans the energy eigenspace. Thus, constructing a generalized density matrix for a non-Hermitian evolution and consequently calculating quantum thermodynamic quantities, such as ergotropy, becomes easier. The generalized density matrix is observed to have a unit trace and contains parameters of the $\pt$symmetric Hamiltonian along with population and coherence terms. The ergotropy is analytically expressed after obtaining the passive state in terms of the modified projectors of the bi-orthonormal basis and the eigenvalues of the generalized density matrix. To understand ergotropy in a $\pt$symmetric open system, a bath was coupled to it in accordance with a Jaynes-Cummings type interaction. Using the $\eta$-pseudo Hermiticity of the $\pt$symmetric Hamiltonian and the $\eta$-unitarity of the total evolution, the generalized system density matrix is obtained by tracing out the radiation degrees of freedom of the bath. Ergotropy was plotted for different initial states, and it was observed that the $\pt$symmetric excited state provided the most extractable work. Using the tunability of non-Hermiticity of the $\pt$symmetric Hamiltonian, different regimes between the normal and the exceptional points are studied. The ergotropy was found to be larger in the Hermitian limit of the $\pt$symmetric Hamiltonian at the normal point. This analysis, thus, paves the way for the realization of a quantum battery using $\pt$symmetric quantum systems.
In the same open system settings, other thermodynamic quantities, namely heat, work, internal energy, entropy production, and von-Neumann entropy, were calculated, and the consistency of the three laws of thermodynamics was verified, thereby confirming that a $\pt$symmetric system is thermodynamically sound. 

We work with a two-level $\pt$symmetric Hamiltonian that can be written in terms of a Hermitian basis by virtue of a characteristic anti-commutation relation, which follows with its complex conjugate. This makes it possible to analyze the states and the eigenvectors along with any subsequent quantum thermodynamical quantities in an extended parameter space ranging from the exceptional point to the Hermitian limit at the normal point. We calculate exact analytical forms of the generalized density matrix and ergotropy, and they show how the Hamiltonian's geometry affects them. The Hermitian basis reveals a deep geometric structure of the eigenspace of the $\eta$-Hermitian Hamiltonian in state space. In~\cite{bose2025PT} we performed an open system treatment of a more general $\pt$symmetric Hamiltonian by coupling the non-Hermitian Hamiltonian with a Hermitian bath and observed different information theoretic measures. In this paper, we calculate basic thermodynamic quantities using the generalized density matrix suitable for the bi-orthonormal basis and check their validity. We calculate ergotropy, heat, work, entropy production, etc., and check the validity of the laws of quantum thermodynamics. This confirms the $\pt$symmetric system as a quantum thermodynamically sound system that is fit to undergo experimentation. Also the form of generalized density matrix we proposed in~\cite{bose2025PT} is testified as a legitimate density matrix capable of representing the dynamics of the $\pt$symmetric Hamiltonian. This type of general non-Markovian open system approach of coupling the non-Hermitian system with a Hermitian bath, coupled in a Jaynes-Cummings type fashion, and the analytical analysis of the generalized density matrix by explicit calculation of the modified projectors is a novel approach. The tunability of the parameter space makes it convenient to extract comparisons of thermodynamic quantities like ergotropy between systems having different degrees of Hermiticity.  Redefining quantum thermodynamical quantities as a consequence of this approach, which satisfies fundamental thermodynamic relations, motivates us to deem this as a valid method to work on further quantum thermodynamics of $\pt$ symmetric or any $\eta$-Hermitian system.
\appendix
\section{Calculation of the generalized density matrix}
To calculate $\rho_G(t)$, we need to calculate the individual projectors using the right eigenvectors and their corresponding left(dual) eigenvectors in Eq.~\eqref{Eneigenvec}.
For the Hamiltonian in Eq.~\eqref{Hpt} the explicit eigenvectors are figured out using Eq.~\eqref{feigenve},
\begin{align}
    \ket{E_\pm^R}&=\frac{1}{2}\left\{ \begin{pmatrix}
        -i\\
        1
    \end{pmatrix}\mp|a|\begin{pmatrix}
        -1\\
        i
    \end{pmatrix} \right\}, \\
    \ket{E_{\pm}^L}&=\frac{1}{2}\left\{ \begin{pmatrix}
        -i\\
        1
    \end{pmatrix}\mp \frac{1}{|a|}\begin{pmatrix}
        -1\\
        i
    \end{pmatrix} \right\},
\end{align}
where $|a|=\left(\frac{f}{1-f}\right)^{\frac{1}{4}}=\sqrt{\frac{s+r}{s-r}}$ using Eq.~\eqref{feigenVa}.
These explicit expressions act as the building blocks of
\begin{align}
    \rho_G(t)=\sum_{ij}\rho^{ij}(t)\ket{E_i^R}\bra{E_j^L},
\end{align}
and the final element-wise expression is obtained as
\begin{widetext}
\begin{align}\label{RhoGt}
    \rho_G^{11}(t)&=\frac{1}{4\sqrt{s^2-r^2}}\left( -i(\sqrt{s+r}+i\sqrt{s-r})^2\rho^{11}(t) -2is\rho^{12}(t) +2is\rho^{21}(t)+i(\sqrt{s+r}-i\sqrt{s-r})^2\rho^{22}(t)\right) \nonumber \\
    \rho_G^{12}(t)&=\frac{1}{4\sqrt{s^2-r^2}}\left( -2s\rho^{11}(t) -(\sqrt{s+r}+i\sqrt{s-r})^2\rho^{12}(t) -(\sqrt{s+r}+i\sqrt{s-r})^2\rho^{21}(t)+2s\rho^{22}(t)\right) \nonumber \\
    \rho_G^{21}(t)&=\frac{1}{4\sqrt{s^2-r^2}}\left( -2s\rho^{11}(t) -(\sqrt{s+r}-i\sqrt{s-r})^2\rho^{12}(t) -(\sqrt{s+r}-i\sqrt{s-r})^2\rho^{21}(t)+2s\rho^{22}(t)\right) \nonumber \\
    \rho_G^{22}(t)&=\frac{1}{4\sqrt{s^2-r^2}}\left( i(\sqrt{s+r}-i\sqrt{s-r})^2\rho^{11}(t) +2is\rho^{12}(t) -2is\rho^{21}(t)-i(\sqrt{s+r}+i\sqrt{s-r})^2\rho^{22}(t)\right). 
\end{align}
\end{widetext}
From this, it can be seen after some algebra that the trace of $\rho_G(t)$ is one, that is, $\rho_G^{11}(t) +  \rho_G^{22}(t) = 1$. Eigenvalues of $\rho_G(t)$, using the above quantities, are
\begin{align}
    \lambda_{\pm}=\frac{1}{2}\pm\frac{1}{2}\sqrt{1-4\left(\rho_G^{11}(t)\rho_G^{22}(t)-\rho_G^{12}(t)\rho_G^{21}(t) \right)},
\end{align}
and are explicitly given in Eq.~\eqref{RhoGeigenvalue}.
An interesting fact that emerges from Eq.~(\ref{RhoGt}) is that the vectorized $\rho_G(t)$ and $\rho(t)$ matrices are connected by a superoperator matrix. Here, the vectorized $\rho(t)$ is a vector in the computational basis made up by the population and coherence terms. The de-vectorized $\rho(t)$ is not a proper density matrix for the $\eta$-pseudo Hermitian, $\pt$symmetric evolution. This superoperator matrix can be assumed to be a higher-dimensional manifestation of the $\eta$ metric that connects the quantum state elements to the vector of a proper density matrix, i.e., a generalized density matrix $\rho_G(t)$ suitable for the $\pt$symmetric space.
and concurrence, for similar setups.
\section{The left eigenvectors}
For real eigenvalues of the $\pt$symmetric $H$, we have,
\begin{align}\label{lefteigenvecs}
    &H\ket{E_n^R}=E_n\ket{E_n^R}.
\end{align}
Taking the complex conjugate and using the $\eta$-Hermiticity
\begin{align}
    \bra{E_n^R}H^{\dagger}=\bra{E_n^R}\eta H\eta^{-1}=E_n\bra{E_n^R}
\end{align}
is obtained and using $\bra{E_n^R}\eta=\bra{E_n^L}$, we have
\begin{align}
    \bra{E_n^L} H=E_n\bra{E_n^L}.
\end{align}
By complex conjugation, we finally obtain
\begin{align}
    H^{\dagger}\ket{E_n^L}=E_n\ket{E_n^L}.
\end{align}
This tells us that the left eigenvectors are the eigenvectors of $H^{\dagger}$.

\section{The $\eta$-adjoint of the unitary operator $U^{\ddagger}$}
Here, we illustrate the
$\eta-$adjoint of the $\eta$-unitary operator. We define $\tilde{\eta}=\eta\otimes I_B$ since $U$ operates on the composite Hilbert space of the system and the bath, and we need to exploit the system Hamiltonian $H$'s $\eta$-Hermiticity, $\eta^{-1}H^{\dagger}\eta=H$. Thus,
\begin{align}
    U^{\ddagger}&=\tilde{\eta}^{-1}U^{\dagger}\tilde{\eta} =\tilde{\eta}^{-1}e^{i\tilde{H}_U^\dagger t}\tilde{\eta}=\tilde{\eta}^{-1}\sum_{n=0}^\infty \frac{(i\tilde{H}_U^\dagger t)^n}{n!}\tilde{\eta} \nonumber \\
    &=\tilde{\eta}^{-1}\left(\sum_{n=0}^\infty \frac{(i t)^n}{n!}(\tilde{H}_U^\dagger)^n \right)\tilde{\eta} \nonumber \\
    &=\sum_{n=0}^\infty \frac{(i t)^n}{n!}\tilde{\eta}^{-1}(\tilde{H}_U^\dagger)^n \tilde{\eta} \nonumber \\
    &=\sum_{n=0}^\infty \frac{(i t)^n}{n!}(\tilde{\eta}^{-1}\tilde{H}_U^\dagger \tilde{\eta})^n  \nonumber \\
    &=\sum_{n=0}^\infty \frac{(i t)^n}{n!}(\tilde{H}_U^\ddagger)^n) \nonumber \\ 
    &=\sum_{n=0}^\infty \frac{(i\tilde{H}_U^\ddagger t)^n}{n!} \nonumber \\ 
    &=\sum_{n=0}^\infty \frac{(i\tilde{H}_U t)^n}{n!} =e^{i\tilde{H}_U t}.
\end{align}
Here, the $\eta$-Hermiticity condition of $\tilde{H}_U$ is used \eqref{etahermiofComposite}.

\section{Redefining Pauli operators in the bi-orthonormal basis}
As $\pt$symmetric system operators, we have used Eqs.~(\ref{Biortho_sigmas}). Here we see how the bi-orthonormal basis vectors build the suitable Pauli $\sigma_G$ operators.
It should be noted that $\sigma_G$'s are not linked with the usual Pauli $\sigma$ matrices with relations like $\sigma_G=\sigma\eta$, i.e., the bi-orthonormal Pauli matrices are not obtained by operating the $\eta$ matrix by means of a matrix product to the $\sigma$ matrices based on the computational basis.
The function of $\eta$ is to modify the projectors of any two-level operators in the bi-orthonormal space by virtue of the equation $\ket{E_n^L}=\eta\ket{E_n^R}$. Let us take $\sigma_G^+$ as an example
\begin{align}
    \sigma_G^+&=\ket{E_+^R}\bra{E_-^L} =\ket{E_+^R}\bra{E_-^R}\eta.
\end{align}
The $\ket{E_+^R}\bra{E_-^R}$ is obtained using the right eigenvectors listed in Eq.~(\ref{Eneigenvec}). The operator $\sigma^+$, on the other hand, is simply
\begin{align}
    \sigma^+&=\ket{0}\bra{1} =\begin{pmatrix}
                1\\
                0
            \end{pmatrix}\begin{pmatrix}
                0 & 1
            \end{pmatrix} =\begin{pmatrix}
                0&1 \\
                0&0
            \end{pmatrix}.
\end{align}
Therefore, it is evident from the above that 
\begin{equation}
    \sigma_G^+\neq\sigma^+\eta.
\end{equation}

The bi-orthonormal Pauli operators follow the commutation relation $[\sigma_G^i, \sigma_G^j] = 2i \, \varepsilon^{ijk} \, \sigma_G^k$.
As an example, we calculate $[\sigma^x_G,\sigma^y_G]$ to show the validity of the above commutation relation.
We have
\begin{align}
    \sigma^x_G\sigma^y_G &=(\ket{E_-^R}\bra{E_+^L}+\ket{E_+^R}\bra{E_-^L} )(i\ket{E_-^R}\bra{E_+^L} \nonumber \\ 
    &-i\ket{E_+^R}\bra{E_-^L} ) \nonumber \\
    &=-i\ket{E_-^R}\bra{E_-^L}+i\ket{E_+^R}\bra{E_+^L},
\end{align}
and
\begin{align}
    \sigma^y_G\sigma^x_G&=(i\ket{E_-^R}\bra{E_+^L}-i\ket{E_+^R}\bra{E_-^L} )(\ket{E_-^R}\bra{E_+^L} \nonumber \\
    &+\ket{E_+^R}\bra{E_-^L} ) \nonumber \\
    &=i\ket{E_-^R}\bra{E_-^L}-i\ket{E_+^R}\bra{E_+^L}.
\end{align}
Subtracting them, we get
\begin{align}
    [\sigma^x_G,\sigma^y_G]&=-2i\ket{E_-^R}\bra{E_-^L}+2i\ket{E_+^R}\bra{E_+^L} \nonumber \\
    &=2i\sigma^z_G,
\end{align}
which verifies the above commutation relation.

\section{Ergotropy for the $\pt$symmetric bi-orthonormal basis}
To define the ergotropy~\cite{Allahverdyan_2004} for the $\pt$symmetric system, we must work with the generalized density matrix $\rho_G(0)=\sum_{ij}\rho_{ij}(0) \ket{E_i^R}\bra{E_j^L}$. For an open system, we can assume $\rho_G(0)$ as the reduced density matrix at time $t=0$ when the interaction with the bath is turned off. We are left with $\rho_G(0)$ to extract work from. We use a time-dependent $\pt$symmetric Hamiltonian $V(t)$ with the boundary condition $V(\tau)=V(0)=0$, making it a cyclic process. The system Hamiltonian that drives the system is the $\pt$symmetric Hamiltonian $H$, rendering the total Hamiltonian as a time-dependent $H_0(t)=H+V(t)$. In Appendix C of~\cite{bose2025PT} we have shown that an evolution driven by an $\eta$-Hermitian $\pt$symmetric Hamiltonian is actually an $\eta$-unitary evolution, i.e., $U^{\ddagger}U=\mathbb{I}$. For an isolated unitary process driven by a time-dependent $\pt$symmetric Hamiltonian, we can generalize the work done on the system as 
\begin{align}\label{workergo}
        dW &=\text{Tr}(\rho_G(t)  \dot{H}_0(t))dt, 
\end{align}
Integrating the above equation, we get
\begin{align}
        W&=\text{Tr}(\rho_G(\tau)H_0(\tau)-\rho_G(0)H_0(0)) \nonumber \\
        &=\text{Tr}[(\rho_G(\tau)-\rho_G(0))H] \nonumber \\
        &=\text{Tr}[\rho_G(\tau)H]-\text{Tr}[\rho_G(0))H] \nonumber \\
        &=E_f-\text{Tr}[\rho_G(0)H]
\end{align}
In~\cite{Allahverdyan_2004}, the ergotropy is defined as $\mathcal{\mathcal{W}}=\max_{U}\{-W(U)\}$, where $E_f$ is the minimum final energy and the maximization is performed over all cyclic unitary operations acting on the system, consistent with the standard definition of ergotropy. Since our system is finite, an isolated unitary evolution would not only be entropy preserving, but it would also preserve the eigenvalues of the generalized density matrix. We must stress that the above quantities are all suitable only for the $\eta$-inner product space specified by the modified eigenvectors of the $\pt$symmetric Hamiltonian.
    To make $E_f=\text{Tr}(\rho_G(\tau)H)$ minimum, we use the variations of $\eta$-unitary evolutions of the initial generalized density matrix $\rho_G(0)$, $\rho_G(\tau)=U\rho_G(0)U^{\ddagger}$. Parameterizing the variation $\delta U=YU$ where Y is an infinitasimal anti-$\eta$-Hermitian matrix, $\delta U^{\ddagger}=-U^{\ddagger}Y$.
    To have the minimum of $E_f$, we must have $\delta E_f=0$, where
    \begin{align}
        \delta E_f =&\text{Tr}\{\delta U\rho_G(0)U^{\ddagger}H+U\rho_G(0)\delta U^{\ddagger}H\} \nonumber \\
        =&\text{Tr}\{Y U\rho_G(0)U^{\ddagger}H-U\rho_G(0) U^{\ddagger}YH\} \nonumber \\
        =&\text{Tr}\{Y\rho_G(\tau)H-\rho_G(\tau)YH\} \nonumber \\
        =&\text{Tr}\{Y[\rho_G(\tau),H]\}.
    \end{align}
The minimization of the final energy $E_f=\text{Tr}(\rho_G(\tau)H)$ implies the commutation relation $[\rho_G(\tau), H]=0$, which implies that $H$ and the final generalized density matrix or the generalized passive state should share an eigenbasis. Since eigenvalues of the generalized density matrices do not change, we can only shuffle them in the energy eigenbasis such that the energy is the lowest. To do that, we put the highest probability (the highest eigenvalue of the state) in the lowest energy state and so on. This creates a generalized passive state,
\begin{align}
    \rho_G(\tau)=\rho^{\mathrm{p}}=\lambda_+\ket{E_+^R}\bra{E_+^L}+\lambda_-\ket{E_-^R}\bra{E_-^L},
\end{align}
where $E_+>E_-$, and $\lambda_+<\lambda_-$ should be the choice for the construction. Having the passive state and using \eqref{workergo}, we can finally define the ergotropy as,
\begin{align}
    \mathcal{W}&=\text{max}(-W)=\text{Tr}[\rho_G(0)H]-E_f \nonumber \\
    &=\text{Tr}[\rho_G(0)H]-\text{Tr}[\rho_G(\tau)H] \nonumber \\
    &=\text{Tr}[\rho_G(0)H]-\text{Tr}[\rho^{\mathrm{p}}H].
\end{align}
This is the expression of ergotropy generalized for an $\eta$-inner product space for an $\eta$-unitary process. In the section that analyzes an open system evolution, we use the reduced generalized density matrix $\rho_G(t)$ as the initial state in the above expression for calculating ergotropy.

\section{Conservation of energy in the bi-orthonormal basis}
The operator for energy is the $\pt$symmetric Hamiltonian~\eqref{Hpt}. If we calculate the change in average energy, we have
\begin{align}\label{energyconser1}
    \frac{d\langle H \rangle}{dt}=\frac{d}{dt} \text{Tr} \big[\rho_G(t) H\big] =\text{Tr}\bigg[\frac{d\rho_G(t)}{dt}   H \bigg].
\end{align}
In our previous work~\cite{bose2025PT}, we have seen how the Schr\"{o}dinger equation for the generalized density matrix $\rho_G(t)=\sum_{ij}\rho^{ij}(t)\ket{E_i^R}\bra{E_i^L}$ as
\begin{align}\label{eq_gen_von_neumann_eq}
    \frac{d}{dt}\rho_G(t)=-i[H,\rho_G(t)].
\end{align}
This is a testimony to the fact that $H$ is the energy operator of the evolution. Applying this in~\eqref{energyconser1}, and using the cyclicity of the trace, we obtain the conservation of energy,
\begin{align}
    \frac{d\langle H \rangle}{dt}=0.
\end{align}

\section{Definition of work transfer for the open $\pt$symmetric Hamiltonian}
In~\cite{Esposito_2010,landi_RevModPhys.93.035008,deffner_book} it is shown that the entropy production of an open system quantifies the irreversibility as the combined effect of the mutual information between the system and the bath and the Kullback-Leibler entropy of the initial and the final state of the bath. When applied to a thermal environment of arbitrary size, the entropy production ($\Sigma$) becomes the combined effect of the change in the von Neumann entropy $\Delta S_S$ of the system and the entropy flow of the system with the environment, characterized by the heat exchanged between the system and the bath
\begin{align}
    \Sigma=\Delta S_S + \beta dQ_B,
\end{align}\\
where the bath is in an equilibrium state $\rho_B(0)=\rho_B^{th}=e^{-\beta H_B}/\text{Tr}(e^{-\beta H_B})$ with an inverse temperature $\beta$.
The heat exchanged is defined as the change in bath energy,
\begin{align}
    dQ_B=\text{Tr}[H_B(\rho_B(t)-\rho_B^{th})].
\end{align}
We already have the change in internal energy of the $\pt$symmetric system as,
\begin{align}
    dU=\text{Tr}[H(\rho_G(t)-\rho_G(0))].
\end{align} 
Now, we define work for a partially $\eta$-unitary evolution described in~\cite{bose2025PT} that includes a $\eta$-Hermitian system Hamiltonian and a Hermitian bath Hamiltonian along with an interaction Hamiltonian where $\eta$-modified system operators are coupled with Hermitian bath operators,
\begin{align}
    \tilde{H}_U&=H+H_B+H_{GB},
\end{align}
where the time independent interaction Hamiltonian $H_{GB}=g(\sigma^+_G\otimes a + \sigma^-_G \otimes a^{\dagger})$.
The total evolution is
\begin{align}
    \rho_{G}(t)=\text{Tr}_B[U\left\{\rho_G(0)\otimes \rho_B(0)\right\}U^{\ddagger}],
\end{align}
where $U^{\ddagger}=(\eta\otimes \mathbb{I}_B)^{-1}U^{\dagger}(\eta\otimes \mathbb{I}_B)=e^{i\tilde{H}_U t}$~\cite{bose2025PT}.
We use the Ehrenfest theorem for expectation values of a  quantum mechanical operator A without explicit time dependence,
\begin{align}
    \frac{d}{dt}\langle A \rangle=-i\langle [A,\tilde{H}_U]\rangle.
\end{align}
In the $\pt$symmetric subspace, the trace is taken in terms of the modified eigenvectors, i.e., $\sum_i\dbra{E_i}.\ket{E_i}$ ~\cite{bose2025PT}. Using the Ehrenfest theorem for $H,H_B$ and $H_{GB}$, we arrive at the relation,
\begin{align}
    \frac{d}{dt}\langle H_{GB} \rangle=-\frac{d}{dt}\langle H \rangle-\frac{d}{dt}\langle H_B \rangle
\end{align}
Integrating this equation and using the above-mentioned definitions of $dU$ and $dQ_B$
\begin{align}
    \langle H_{GB} \rangle|^t_0&=-\langle H \rangle|^t_0-\langle H_B \rangle|^t_0, \nonumber \\
    -\text{Tr}[(\rho_{GB}(t)-\rho_{GB}(0))H_{GB}]&=\text{Tr}[(\rho_{G}(t)-\rho_{G}(0))H] \nonumber \\+&\text{Tr}[(\rho_{B}(t)-\rho_{B}(0))H_B], \nonumber \\
     \text{Tr}[(\rho_{GB}(0)-\rho_{GB}(t))H_{GB}]&= dU +dQ_B.
\end{align}
Now, to maintain the validity of the first law of thermodynamics
\begin{align}
    dU=dW-dQ_B,
\end{align}
work is defined as,
\begin{align}
    dW=\text{Tr}[(\rho_{GB}(0)-\rho_{GB}(t))H_{GB}].
\end{align}

\bibliographystyle{apsrev4-1}
\bibliography{reference}

@article{Bender2002,
  title = {Complex Extension of Quantum Mechanics},
  volume = {89},
  ISSN = {1079-7114},
  url = {http://dx.doi.org/10.1103/PhysRevLett.89.270401},
  DOI = {10.1103/physrevlett.89.270401},
  number = {27},
  journal = {Physical Review Letters},
  publisher = {American Physical Society (APS)},
  author = {Bender,  Carl M. and Brody,  Dorje C. and Jones,  Hugh F.},
  year = {2002},
  month = dec 
}

@article{Brody_2014,
doi = {10.1088/1751-8113/47/3/035305},
url = {https://doi.org/10.1088/1751-8113/47/3/035305},
year = {2013},
month = {dec},
publisher = {IOP Publishing},
volume = {47},
number = {3},
pages = {035305},
author = {Brody, Dorje C},
title = {Biorthogonal quantum mechanics},
journal = {Journal of Physics A: Mathematical and Theoretical},
}

@article{Bender_2004,
doi = {10.1088/0305-4470/37/43/009},
url = {https://dx.doi.org/10.1088/0305-4470/37/43/009},
year = {2004},
month = {oct},
publisher = {},
volume = {37},
number = {43},
pages = {10139},
author = {Carl M Bender and Joachim Brod and André Refig and Moretz E Reuter},
title = {The $\mathcal{C}$ operator in $\mathcal{PT}$-symmetric quantum theories},
journal = {Journal of Physics A: Mathematical and General},

}

@article{Bender1,
  title = {Real Spectra in Non-Hermitian Hamiltonians Having $\mathsc{P}\mathsc{T}$ Symmetry},
  author = {Bender, Carl M. and Boettcher, Stefan},
  journal = {Phys. Rev. Lett.},
  volume = {80},
  issue = {24},
  pages = {5243--5246},
  numpages = {0},
  year = {1998},
  month = {Jun},
  publisher = {American Physical Society},
  doi = {10.1103/PhysRevLett.80.5243},
  url = {https://link.aps.org/doi/10.1103/PhysRevLett.80.5243}
}

@article{Bender2_1999,
  title = {PT-symmetric quantum mechanics},
  volume = {40},
  ISSN = {1089-7658},
  url = {http://dx.doi.org/10.1063/1.532860},
  DOI = {10.1063/1.532860},
  number = {5},
  journal = {Journal of Mathematical Physics},
  publisher = {AIP Publishing},
  author = {Bender,  Carl M. and Boettcher,  Stefan and Meisinger,  Peter N.},
  year = {1999},
  month = may,
  pages = {2201–2229}
}

@article{Bender_4_2003,
doi = {10.1088/0305-4470/36/24/314},
url = {https://dx.doi.org/10.1088/0305-4470/36/24/314},
year = {2003},
month = {jun},
publisher = {},
volume = {36},
number = {24},
pages = {6791},
author = {Carl M Bender and Peter N Meisinger and Qinghai Wang},
title = {Finite-dimensional ����-symmetric Hamiltonians},
journal = {Journal of Physics A: Mathematical and General},
}

@article{Bender_5_2005,
  title = {Introduction to ����-symmetric quantum theory},
  volume = {46},
  ISSN = {1366-5812},
  url = {http://dx.doi.org/10.1080/00107500072632},
  DOI = {10.1080/00107500072632},
  number = {4},
  journal = {Contemporary Physics},
  publisher = {Informa UK Limited},
  author = {Bender,  Carl M},
  year = {2005},
  month = jul,
  pages = {277–292}
}

@article{Bender_6_2007,
doi = {10.1088/0034-4885/70/6/R03},
url = {https://dx.doi.org/10.1088/0034-4885/70/6/R03},
year = {2007},
month = {may},
publisher = {},
volume = {70},
number = {6},
pages = {947},
author = {Bender, Carl M},
title = {Making sense of non-Hermitian Hamiltonians},
journal = {Reports on Progress in Physics},

}

@article{Mostafazadeh_1_2002,
  title = {Pseudo-Hermiticity versus PT-symmetry. II. A complete characterization of non-Hermitian Hamiltonians with a real spectrum},
  volume = {43},
  ISSN = {1089-7658},
  url = {http://dx.doi.org/10.1063/1.1461427},
  DOI = {10.1063/1.1461427},
  number = {5},
  journal = {Journal of Mathematical Physics},
  publisher = {AIP Publishing},
  author = {Mostafazadeh,  Ali},
  year = {2002},
  month = may,
  pages = {2814–2816}
}

@article{Mostafazadeh_2_2002,
  title = {Pseudo-Hermiticity versus PT symmetry: The necessary condition for the reality of the spectrum of a non-Hermitian Hamiltonian},
  volume = {43},
  ISSN = {1089-7658},
  url = {http://dx.doi.org/10.1063/1.1418246},
  DOI = {10.1063/1.1418246},
  number = {1},
  journal = {Journal of Mathematical Physics},
  publisher = {AIP Publishing},
  author = {Mostafazadeh,  Ali},
  year = {2002},
  month = jan,
  pages = {205–214}
}

@article{Mostafazadeh_3_2002,
  title = {Pseudo-Hermiticity versus PT-symmetry III: Equivalence of pseudo-Hermiticity and the presence of antilinear symmetries},
  volume = {43},
  ISSN = {1089-7658},
  url = {http://dx.doi.org/10.1063/1.1489072},
  DOI = {10.1063/1.1489072},
  number = {8},
  journal = {Journal of Mathematical Physics},
  publisher = {AIP Publishing},
  author = {Mostafazadeh,  Ali},
  year = {2002},
  month = aug,
  pages = {3944–3951}
}

@article{Mostafazadeh_4_2004,
doi = {10.1088/0305-4470/37/48/009},
url = {https://dx.doi.org/10.1088/0305-4470/37/48/009},
year = {2004},
month = {nov},
publisher = {},
volume = {37},
number = {48},
pages = {11645},
author = {Mostafazadeh, Ali and Batal, Ahmet},
title = {Physical aspects of pseudo-Hermitian and PT-symmetric quantum mechanics},
journal = {Journal of Physics A: Mathematical and General},

}

@article{Mostaf_5_doi:10.1142/S0219887810004816,
author = {MOSTAFAZADEH, ALI},
title = {PSEUDO-HERMITIAN REPRESENTATION OF QUANTUM MECHANICS},
journal = {International Journal of Geometric Methods in Modern Physics},
volume = {07},
number = {07},
pages = {1191-1306},
year = {2010},
doi = {10.1142/S0219887810004816},

URL = { 
    
        https://doi.org/10.1142/S0219887810004816
},
eprint = { 
        https://doi.org/10.1142/S0219887810004816

}
,
}

@article{Optics_Rter2010,
  title = {Observation of parity–time symmetry in optics},
  volume = {6},
  ISSN = {1745-2481},
  url = {http://dx.doi.org/10.1038/nphys1515},
  DOI = {10.1038/nphys1515},
  number = {3},
  journal = {Nature Physics},
  publisher = {Springer Science and Business Media LLC},
  author = {R\"{u}ter,  Christian E. and Makris,  Konstantinos G. and El-Ganainy,  Ramy and Christodoulides,  Demetrios N. and Segev,  Mordechai and Kip,  Detlef},
  year = {2010},
  month = jan,
  pages = {192–195}
}

@article{Electronics_PhysRevA.84.040101,
  title = {Experimental study of active LRC circuits with $\mathcal{PT}$ symmetries},
  author = {Schindler, Joseph and Li, Ang and Zheng, Mei C. and Ellis, F. M. and Kottos, Tsampikos},
  journal = {Phys. Rev. A},
  volume = {84},
  issue = {4},
  pages = {040101},
  numpages = {5},
  year = {2011},
  month = {Oct},
  publisher = {American Physical Society},
  doi = {10.1103/PhysRevA.84.040101},
  url = {https://link.aps.org/doi/10.1103/PhysRevA.84.040101}
}

@article{microwaves_PhysRevLett.108.024101,
  title = {$\mathsc{P}\mathsc{T}$ Symmetry and Spontaneous Symmetry Breaking in a Microwave Billiard},
  author = {Bittner, S. and Dietz, B. and G\"unther, U. and Harney, H. L. and Miski-Oglu, M. and Richter, A. and Sch\"afer, F.},
  journal = {Phys. Rev. Lett.},
  volume = {108},
  issue = {2},
  pages = {024101},
  numpages = {5},
  year = {2012},
  month = {Jan},
  publisher = {American Physical Society},
  doi = {10.1103/PhysRevLett.108.024101},
  url = {https://link.aps.org/doi/10.1103/PhysRevLett.108.024101}
}

@article{mechanical_Bender2013,
  title = {Observation of PT phase transition in a simple mechanical system},
  volume = {81},
  ISSN = {1943-2909},
  url = {http://dx.doi.org/10.1119/1.4789549},
  DOI = {10.1119/1.4789549},
  number = {3},
  journal = {American Journal of Physics},
  publisher = {American Association of Physics Teachers (AAPT)},
  author = {Bender,  Carl M. and Berntson,  Bjorn K. and Parker,  David and Samuel,  E.},
  year = {2013},
  month = feb,
  pages = {173–179}
}

@article{acoustic_Fleury2015,
  title = {An invisible acoustic sensor based on parity-time symmetry},
  volume = {6},
  ISSN = {2041-1723},
  url = {http://dx.doi.org/10.1038/ncomms6905},
  DOI = {10.1038/ncomms6905},
  number = {1},
  journal = {Nature Communications},
  publisher = {Springer Science and Business Media LLC},
  author = {Fleury,  Romain and Sounas,  Dimitrios and Alù,  Andrea},
  year = {2015},
  month = jan 
}

@article{atomic_1_Baker1984,
  title = {Non-Hermitian quantum theory of multiphoton ionization},
  volume = {30},
  ISSN = {0556-2791},
  url = {http://dx.doi.org/10.1103/PhysRevA.30.773},
  DOI = {10.1103/physreva.30.773},
  number = {2},
  journal = {Physical Review A},
  publisher = {American Physical Society (APS)},
  author = {Baker,  Howard C.},
  year = {1984},
  month = aug,
  pages = {773–793}
}

@article{atomic_1_PhysRevLett.110.083604,
  title = {$\mathcal{P}\mathcal{T}$ Symmetry with a System of Three-Level Atoms},
  author = {Hang, Chao and Huang, Guoxiang and Konotop, Vladimir V.},
  journal = {Phys. Rev. Lett.},
  volume = {110},
  issue = {8},
  pages = {083604},
  numpages = {5},
  year = {2013},
  month = {Feb},
  publisher = {American Physical Society},
  doi = {10.1103/PhysRevLett.110.083604},
  url = {https://link.aps.org/doi/10.1103/PhysRevLett.110.083604}
}

@article{atomic_2_PhysRevLett.117.123601,
  title = {Observation of Parity-Time Symmetry in Optically Induced Atomic Lattices},
  author = {Zhang, Zhaoyang and Zhang, Yiqi and Sheng, Jiteng and Yang, Liu and Miri, Mohammad-Ali and Christodoulides, Demetrios N. and He, Bing and Zhang, Yanpeng and Xiao, Min},
  journal = {Phys. Rev. Lett.},
  volume = {117},
  issue = {12},
  pages = {123601},
  numpages = {5},
  year = {2016},
  month = {Sep},
  publisher = {American Physical Society},
  doi = {10.1103/PhysRevLett.117.123601},
  url = {https://link.aps.org/doi/10.1103/PhysRevLett.117.123601}
}

@article{
singlespin_doi:10.1126/science.aaw8205,
author = {Yang Wu  and Wenquan Liu  and Jianpei Geng  and Xingrui Song  and Xiangyu Ye  and Chang-Kui Duan  and Xing Rong  and Jiangfeng Du },
title = {Observation of parity-time symmetry breaking in a single-spin system},
journal = {Science},
volume = {364},
number = {6443},
pages = {878-880},
year = {2019},
doi = {10.1126/science.aaw8205},
URL = {https://www.science.org/doi/abs/10.1126/science.aaw8205},
eprint = {https://www.science.org/doi/pdf/10.1126/science.aaw8205},
}

@article{Scolarici2006,
  title = {Time evolution of non-Hermitian quantum systems and generalized master equations},
  volume = {56},
  ISSN = {1572-9486},
  url = {http://dx.doi.org/10.1007/s10582-006-0389-7},
  DOI = {10.1007/s10582-006-0389-7},
  number = {9},
  journal = {Czechoslovak Journal of Physics},
  publisher = {Springer Science and Business Media LLC},
  author = {Scolarici,  G. and Solombrino,  L.},
  year = {2006},
  month = sep,
  pages = {935–941}
}

@article{Scolarici2007,
  title = {Complex Projection of Quasianti-Hermitian Quaternionic Hamiltonian Dynamics},
  ISSN = {1815-0659},
  url = {http://dx.doi.org/10.3842/SIGMA.2007.088},
  DOI = {10.3842/sigma.2007.088},
  journal = {Symmetry,  Integrability and Geometry: Methods and Applications},
  publisher = {SIGMA (Symmetry,  Integrability and Geometry: Methods and Application)},
  author = {Scolarici,  Giuseppe},
  year = {2007},
  month = sep 
}

@article{TommyOhlsson_densitymatrixforma_PhysRevA.103.022218,
  title = {Density-matrix formalism for $\mathcal{PT}$-symmetric non-Hermitian Hamiltonians with the Lindblad equation},
  author = {Ohlsson, Tommy and Zhou, Shun},
  journal = {Phys. Rev. A},
  volume = {103},
  issue = {2},
  pages = {022218},
  numpages = {20},
  year = {2021},
  month = {Feb},
  publisher = {American Physical Society},
  doi = {10.1103/PhysRevA.103.022218},
  url = {https://link.aps.org/doi/10.1103/PhysRevA.103.022218}
}

@article{Alexandre2018PRD,
  author = {Alexandre, J. and Ellis, J. and Millington, P. and Seynaeve, D.},
  title = {Spontaneous symmetry breaking and the Goldstone theorem in non-Hermitian field theories},
  journal = {Physical Review D},
  volume = {98},
  pages = {125003},
  year = {2018},
  doi = {10.1103/PhysRevD.98.125003}
}

@article{neutmassgen_Alexandre2015,
  title = {Non-Hermitian extension of gauge theories and implications for neutrino physics},
  volume = {2015},
  ISSN = {1029-8479},
  url = {http://dx.doi.org/10.1007/JHEP11(2015)111},
  DOI = {10.1007/jhep11(2015)111},
  number = {11},
  journal = {Journal of High Energy Physics},
  publisher = {Springer Science and Business Media LLC},
  author = {Alexandre,  Jean and Bender,  Carl M. and Millington,  Peter},
  year = {2015},
  month = nov 
}

@article{neutosci1_PhysRevD.89.125014,
  title = {Relativistic non-Hermitian quantum mechanics},
  author = {Jones-Smith, Katherine and Mathur, Harsh},
  journal = {Phys. Rev. D},
  volume = {89},
  issue = {12},
  pages = {125014},
  numpages = {5},
  year = {2014},
  month = {Jun},
  publisher = {American Physical Society},
  doi = {10.1103/PhysRevD.89.125014},
  url = {https://link.aps.org/doi/10.1103/PhysRevD.89.125014}
}

@article{neutosci_2_Ohlsson2016,
  title = {Non-Hermitian neutrino oscillations in matter with PT symmetric Hamiltonians},
  volume = {113},
  ISSN = {1286-4854},
  url = {http://dx.doi.org/10.1209/0295-5075/113/61001},
  DOI = {10.1209/0295-5075/113/61001},
  number = {6},
  journal = {EPL (Europhysics Letters)},
  publisher = {IOP Publishing},
  author = {Ohlsson,  Tommy},
  year = {2016},
  month = mar,
  pages = {61001}
}

@article{Yukawa_Alexandre2017,
  title = {Light neutrino masses from a non-Hermitian Yukawa theory},
  volume = {873},
  ISSN = {1742-6596},
  url = {http://dx.doi.org/10.1088/1742-6596/873/1/012047},
  DOI = {10.1088/1742-6596/873/1/012047},
  journal = {Journal of Physics: Conference Series},
  publisher = {IOP Publishing},
  author = {Alexandre,  J and Bender,  C M and Millington,  P},
  year = {2017},
  month = jul,
  pages = {012047}
}

@article{Goldstone_PhysRevD.98.045001,
  title = {Spontaneous symmetry breaking and the Goldstone theorem in non-Hermitian field theories},
  author = {Alexandre, Jean and Ellis, John and Millington, Peter and Seynaeve, Dries},
  journal = {Phys. Rev. D},
  volume = {98},
  issue = {4},
  pages = {045001},
  numpages = {11},
  year = {2018},
  month = {Aug},
  publisher = {American Physical Society},
  doi = {10.1103/PhysRevD.98.045001},
  url = {https://link.aps.org/doi/10.1103/PhysRevD.98.045001}
}

@article{Brout_higgs_PhysRevD.99.045006,
  title = {Goldstone bosons and the Englert-Brout-Higgs mechanism in non-Hermitian theories},
  author = {Mannheim, Philip D.},
  journal = {Phys. Rev. D},
  volume = {99},
  issue = {4},
  pages = {045006},
  numpages = {19},
  year = {2019},
  month = {Feb},
  publisher = {American Physical Society},
  doi = {10.1103/PhysRevD.99.045006},
  url = {https://link.aps.org/doi/10.1103/PhysRevD.99.045006}
}

@article{brout_higgs_2_PhysRevD.99.075024,
  title = {Gauge invariance and the Englert-Brout-Higgs mechanism in non-Hermitian field theories},
  author = {Alexandre, Jean and Ellis, John and Millington, Peter and Seynaeve, Dries},
  journal = {Phys. Rev. D},
  volume = {99},
  issue = {7},
  pages = {075024},
  numpages = {9},
  year = {2019},
  month = {Apr},
  publisher = {American Physical Society},
  doi = {10.1103/PhysRevD.99.075024},
  url = {https://link.aps.org/doi/10.1103/PhysRevD.99.075024}
}

@book{Breuer2007,
  title = {The Theory of Open Quantum Systems},
  ISBN = {9780191706349},
  url = {http://dx.doi.org/10.1093/acprof:oso/9780199213900.001.0001},
  DOI = {10.1093/acprof:oso/9780199213900.001.0001},
  publisher = {Oxford University PressOxford},
  author = {Breuer,  Heinz-Peter and Petruccione,  Francesco},
  year = {2007},
  month = jan 
}

@book{Banerjee2018,
  title = {Open Quantum Systems: Dynamics of Nonclassical Evolution},
  ISBN = {9789811331824},
  ISSN = {2366-8857},
  url = {http://dx.doi.org/10.1007/978-981-13-3182-4},
  DOI = {10.1007/978-981-13-3182-4},
  journal = {Texts and Readings in Physical Sciences},
  publisher = {Springer Singapore},
  author = {Banerjee,  Subhashish},
  year = {2018}
}

@book{Weiss2011,
  title = {Quantum Dissipative Systems},
  ISBN = {9789814374927},
  url = {http://dx.doi.org/10.1142/8334},
  DOI = {10.1142/8334},
  publisher = {WORLD SCIENTIFIC},
  author = {Weiss,  Ulrich},
  year = {2011},
  month = nov 
}

@article{GKLSpaper,
    author = {Gorini, Vittorio and Kossakowski, Andrzej and Sudarshan, E. C. G.},
    title = "{Completely positive dynamical semigroups of N‐level systems}",
    journal = {Journal of Mathematical Physics},
    volume = {17},
    number = {5},
    pages = {821-825},
    year = {1976},
    month = {05},
    issn = {0022-2488},
    doi = {10.1063/1.522979},
    url = {https://doi.org/10.1063/1.522979},
    eprint = {https://pubs.aip.org/aip/jmp/article-pdf/17/5/821/19090720/821\_1\_online.pdf},
}

@Article{Utagi2020,
author={Utagi, Shrikant
and Srikanth, R.
and Banerjee, Subhashish},
title={Temporal self-similarity of quantum dynamical maps as a concept of memorylessness},
journal={Scientific Reports},
year={2020},
month={Sep},
day={14},
volume={10},
number={1},
pages={15049},
issn={2045-2322},
doi={10.1038/s41598-020-72211-3},
url={https://doi.org/10.1038/s41598-020-72211-3}
}

@article{banerjeepetrucione,
author = {Kumar, N. Pradeep and Banerjee, Subhashish and Srikanth, R. and Jagadish, Vinayak and Petruccione, Francesco},
title = {Non-Markovian Evolution: a Quantum Walk Perspective},
journal = {Open Systems \& Information Dynamics},
volume = {25},
number = {03},
pages = {1850014},
year = {2018},
doi = {10.1142/S1230161218500142},
URL = {https://doi.org/10.1142/S1230161218500142},
eprint = {https://doi.org/10.1142/S1230161218500142}
}

@article{Vacchini_2011,
doi = {10.1088/1367-2630/13/9/093004},
url = {https://dx.doi.org/10.1088/1367-2630/13/9/093004},
year = {2011},
month = {sep},
publisher = {},
volume = {13},
number = {9},
pages = {093004},
author = {Bassano Vacchini and Andrea Smirne and Elsi-Mari Laine and Jyrki Piilo and Heinz-Peter Breuer},
title = {Markovianity and non-Markovianity in quantum and classical systems},
journal = {New Journal of Physics}
}

@article{Tiwari_2023,
author = {Tiwari, Devvrat and Paulson, Kavalambramalil G. and Banerjee, Subhashish},
title = {Quantum Correlations and Speed Limit of Central Spin Systems},
journal = {Annalen der Physik},
volume = {535},
number = {2},
pages = {2200452},
doi = {https://doi.org/10.1002/andp.202200452},
url = {https://onlinelibrary.wiley.com/doi/abs/10.1002/andp.202200452},
year = {2023}
}

@article{Lindblad1976,
  title = {On the generators of quantum dynamical semigroups},
  volume = {48},
  ISSN = {1432-0916},
  url = {http://dx.doi.org/10.1007/BF01608499},
  DOI = {10.1007/bf01608499},
  number = {2},
  journal = {Communications in Mathematical Physics},
  publisher = {Springer Science and Business Media LLC},
  author = {Lindblad,  G.},
  year = {1976},
  month = jun,
  pages = {119–130}
}

@article{Hall_2014,
  title = {Canonical form of master equations and characterization of non-Markovianity},
  author = {Hall, Michael J. W. and Cresser, James D. and Li, Li and Andersson, Erika},
  journal = {Phys. Rev. A},
  volume = {89},
  issue = {4},
  pages = {042120},
  numpages = {11},
  year = {2014},
  month = {Apr},
  publisher = {American Physical Society},
  doi = {10.1103/PhysRevA.89.042120},
  url = {https://link.aps.org/doi/10.1103/PhysRevA.89.042120}
}

@article{Rivas_2014,
doi = {10.1088/0034-4885/77/9/094001},
url = {https://dx.doi.org/10.1088/0034-4885/77/9/094001},
year = {2014},
month = {aug},
publisher = {IOP Publishing},
volume = {77},
number = {9},
pages = {094001},
author = {Ángel Rivas and Susana F Huelga and Martin B Plenio},
title = {Quantum non-Markovianity: characterization, quantification and detection},
journal = {Reports on Progress in Physics}
}

@article{RevModPhys.88.021002,
  title = {Colloquium: Non-Markovian dynamics in open quantum systems},
  author = {Breuer, Heinz-Peter and Laine, Elsi-Mari and Piilo, Jyrki and Vacchini, Bassano},
  journal = {Rev. Mod. Phys.},
  volume = {88},
  issue = {2},
  pages = {021002},
  numpages = {24},
  year = {2016},
  month = {Apr},
  publisher = {American Physical Society},
  doi = {10.1103/RevModPhys.88.021002},
  url = {https://link.aps.org/doi/10.1103/RevModPhys.88.021002}
}

@article{CHRUSCINSKI20221,
title = {Dynamical maps beyond Markovian regime},
journal = {Physics Reports},
volume = {992},
pages = {1-85},
year = {2022},
note = {Dynamical maps beyond Markovian regime},
issn = {0370-1573},
doi = {https://doi.org/10.1016/j.physrep.2022.09.003},
url = {https://www.sciencedirect.com/science/article/pii/S0370157322003428},
author = {Dariusz Chruściński}
}

@article{tiwari2024strong,
    author = {Tiwari, Devvrat and Bose, Baibhab and Banerjee, Subhashish},
    title = {Strong coupling non-Markovian quantum thermodynamics of a finite-bath system},
    journal = {The Journal of Chemical Physics},
    volume = {162},
    number = {11},
    pages = {114104},
    year = {2025},
    month = {03},
    issn = {0021-9606},
    doi = {10.1063/5.0254029},
    url = {https://doi.org/10.1063/5.0254029}
}

@misc{kading2025,
      title={Density matrices in quantum field theory: Non-Markovianity, path integrals and master equations}, 
      author={Christian Käding and Mario Pitschmann},
      year={2025},
      eprint={2503.08567},
      archivePrefix={arXiv},
      primaryClass={hep-th},
      url={https://arxiv.org/abs/2503.08567}, 
}

@ARTICLE{kleefeld_2009arXiv0906.1011K,
       author = {{Kleefeld}, F.},
        title = "{The construction of a general inner product in non-Hermitian quantum theory and some explanation for the nonuniqueness of the C operator in PT quantum mechanics}",
      journal = {arXiv e-prints},
         year = 2009,
        month = jun,
          eid = {arXiv:0906.1011},
        pages = {arXiv:0906.1011},
          doi = {10.48550/arXiv.0906.1011},
archivePrefix = {arXiv},
       eprint = {0906.1011},
 primaryClass = {hep-th},
       adsurl = {https://ui.adsabs.harvard.edu/abs/2009arXiv0906.1011K}
}

@article{Mostafazadeh_2002,
    author = {Mostafazadeh, Ali},
    title = {Pseudo-Hermiticity versus PT symmetry: The necessary condition for the reality of the spectrum of a non-Hermitian Hamiltonian},
    journal = {Journal of Mathematical Physics},
    volume = {43},
    number = {1},
    pages = {205-214},
    year = {2002},
    month = {01},
    issn = {0022-2488},
    doi = {10.1063/1.1418246},
    url = {https://doi.org/10.1063/1.1418246},
    eprint = {https://pubs.aip.org/aip/jmp/article-pdf/43/1/205/19019524/205\_1\_online.pdf},
}

@article{Javid_2019,
  title = {Quantum Zeno effect and nonclassicality in a $\mathcal{PT}$-symmetric system of coupled cavities},
  author = {Naikoo, Javid and Thapliyal, Kishore and Banerjee, Subhashish and Pathak, Anirban},
  journal = {Phys. Rev. A},
  volume = {99},
  issue = {2},
  pages = {023820},
  numpages = {9},
  year = {2019},
  month = {Feb},
  publisher = {American Physical Society},
  doi = {10.1103/PhysRevA.99.023820},
  url = {https://link.aps.org/doi/10.1103/PhysRevA.99.023820}
}

@article{Agarwal_2012,
  title = {Spontaneous generation of photons in transmission of quantum fields in $PT$-symmetric optical systems},
  author = {Agarwal, G. S. and Qu, Kenan},
  journal = {Phys. Rev. A},
  volume = {85},
  issue = {3},
  pages = {031802},
  numpages = {4},
  year = {2012},
  month = {Mar},
  publisher = {American Physical Society},
  doi = {10.1103/PhysRevA.85.031802},
  url = {https://link.aps.org/doi/10.1103/PhysRevA.85.031802}
}

@article{Nori_2019,
  title = {Quantum exceptional points of non-Hermitian Hamiltonians and Liouvillians: The effects of quantum jumps},
  author = {Minganti, Fabrizio and Miranowicz, Adam and Chhajlany, Ravindra W. and Nori, Franco},
  journal = {Phys. Rev. A},
  volume = {100},
  issue = {6},
  pages = {062131},
  numpages = {17},
  year = {2019},
  month = {Dec},
  publisher = {American Physical Society},
  doi = {10.1103/PhysRevA.100.062131},
  url = {https://link.aps.org/doi/10.1103/PhysRevA.100.062131}
}

@article{Badhani_2024,
doi = {10.1088/1402-4896/ad753f},
url = {https://dx.doi.org/10.1088/1402-4896/ad753f},
year = {2024},
month = {sep},
publisher = {IOP Publishing},
volume = {99},
number = {10},
pages = {105112},
author = {Badhani, Himanshu and Banerjee, Subhashish and Chandrashekar, C M},
title = {Non-Hermitian quantum walks and non-Markovianity: the coin-position interaction},
journal = {Physica Scripta}
}

@article{Naikoo_2021,
doi = {10.1088/1751-8121/ac0546},
url = {https://dx.doi.org/10.1088/1751-8121/ac0546},
year = {2021},
month = {jun},
publisher = {IOP Publishing},
volume = {54},
number = {27},
pages = {275303},
author = {Naikoo, Javid and Kumari, Swati and Banerjee, Subhashish and Pan, A K},
title = {\mathcal{P}\mathcal{T} symmetric evolution, coherence and violation of Leggett–Garg inequalities},
journal = {Journal of Physics A: Mathematical and Theoretical}
}

@article{Javid_SB_2019,
  title = {Interplay between nonclassicality and $\mathcal{PT}$ symmetry in an effective two-level system with open system effects},
  author = {Naikoo, Javid and Banerjee, Subhashish and Pathak, Anirban},
  journal = {Phys. Rev. A},
  volume = {100},
  issue = {2},
  pages = {023836},
  numpages = {11},
  year = {2019},
  month = {Aug},
  publisher = {American Physical Society},
  doi = {10.1103/PhysRevA.100.023836},
  url = {https://link.aps.org/doi/10.1103/PhysRevA.100.023836}
}

@article{Das_Bhasin_2025,
doi = {10.1088/1751-8121/adbac6},
url = {https://dx.doi.org/10.1088/1751-8121/adbac6},
year = {2025},
month = {mar},
publisher = {IOP Publishing},
volume = {58},
number = {12},
pages = {125303},
author = {Bhasin, Priyanshi and Das, Tanmoy},
title = {A Hermitian bypass to the non-Hermitian quantum theory},
journal = {Journal of Physics A: Mathematical and Theoretical}
}

@article{Asok_das_2010,
    author = {Das, Ashok and Greenwood, L.},
    title = {An alternative construction of the positive inner product for pseudo-Hermitian Hamiltonians: Examples},
    journal = {Journal of Mathematical Physics},
    volume = {51},
    number = {4},
    pages = {042103},
    year = {2010},
    month = {04},
    issn = {0022-2488},
    doi = {10.1063/1.3373551},
    url = {https://doi.org/10.1063/1.3373551}
}

@article{Kumari_2022,
doi = {10.1088/1751-8121/ac5dae},
url = {https://dx.doi.org/10.1088/1751-8121/ac5dae},
year = {2022},
month = {apr},
publisher = {IOP Publishing},
volume = {55},
number = {18},
pages = {185302},
author = {Kumari, Asmita and Sen, Ujjwal},
title = {Local preservation of no-signaling in multiparty PT-symmetric evolutions},
journal = {Journal of Physics A: Mathematical and Theoretical}
}

@article{7t26_13_PhysRevE.97.062108,
  title = {Thermodynamics of non-Markovian reservoirs and heat engines},
  author = {Thomas, George and Siddharth, Nana and Banerjee, Subhashish and Ghosh, Sibasish},
  journal = {Phys. Rev. E},
  volume = {97},
  issue = {6},
  pages = {062108},
  numpages = {8},
  year = {2018},
  month = {Jun},
  publisher = {American Physical Society},
  doi = {10.1103/PhysRevE.97.062108},
  url = {https://link.aps.org/doi/10.1103/PhysRevE.97.062108}
}

@article{Esposito_2010,
doi = {10.1088/1367-2630/12/1/013013},
url = {https://dx.doi.org/10.1088/1367-2630/12/1/013013},
year = {2010},
month = {jan},
publisher = {},
volume = {12},
number = {1},
pages = {013013},
author = {Massimiliano Esposito and Katja Lindenberg and Christian Van den Broeck},
title = {Entropy production as correlation between system and reservoir},
journal = {New Journal of Physics}
}

@article{landi_RevModPhys.93.035008,
  title = {Irreversible entropy production: From classical to quantum},
  author = {Landi, Gabriel T. and Paternostro, Mauro},
  journal = {Rev. Mod. Phys.},
  volume = {93},
  issue = {3},
  pages = {035008},
  numpages = {58},
  year = {2021},
  month = {Sep},
  publisher = {American Physical Society},
  doi = {10.1103/RevModPhys.93.035008},
  url = {https://link.aps.org/doi/10.1103/RevModPhys.93.035008}
}

@article{full_counting_paper,
  title = {Switching the function of the quantum Otto cycle in non-Markovian dynamics: Heat engine, heater, and heat pump},
  author = {Ishizaki, Miku and Hatano, Naomichi and Tajima, Hiroyasu},
  journal = {Phys. Rev. Res.},
  volume = {5},
  issue = {2},
  pages = {023066},
  numpages = {14},
  year = {2023},
  month = {Apr},
  publisher = {American Physical Society},
  doi = {10.1103/PhysRevResearch.5.023066},
  url = {https://link.aps.org/doi/10.1103/PhysRevResearch.5.023066}
}

@book{deffner_book,
author = {Deffner, Sebastian and Campbell, Steve},
title = {Quantum Thermodynamics},
publisher = {Morgan \& Claypool Publishers},
year = {2019},
series = {2053-2571},
isbn = {978-1-64327-658-8},
url = {https://dx.doi.org/10.1088/2053-2571/ab21c6},
doi = {10.1088/2053-2571/ab21c6}
}

@article{vega_alonso,
  title = {Dynamics of non-Markovian open quantum systems},
  author = {de Vega, In\'es and Alonso, Daniel},
  journal = {Rev. Mod. Phys.},
  volume = {89},
  issue = {1},
  pages = {015001},
  numpages = {58},
  year = {2017},
  month = {Jan},
  publisher = {American Physical Society},
  doi = {10.1103/RevModPhys.89.015001},
  url = {https://link.aps.org/doi/10.1103/RevModPhys.89.015001}
}

@article{Strasberg_2016,
doi = {10.1088/1367-2630/18/7/073007},
url = {https://dx.doi.org/10.1088/1367-2630/18/7/073007},
year = {2016},
month = {jul},
publisher = {IOP Publishing},
volume = {18},
number = {7},
pages = {073007},
author = {Philipp Strasberg and Gernot Schaller and Neill Lambert and Tobias Brandes},
title = {Nonequilibrium thermodynamics in the strong coupling and non-Markovian regime based on a reaction coordinate mapping},
journal = {New Journal of Physics}
}

@article{Zhang_2022,
  title = {Strong-coupling quantum thermodynamics far from equilibrium: Non-Markovian transient quantum heat and work},
  author = {Huang, Wei-Ming and Zhang, Wei-Min},
  journal = {Phys. Rev. A},
  volume = {106},
  issue = {3},
  pages = {032607},
  numpages = {13},
  year = {2022},
  month = {Sep},
  publisher = {American Physical Society},
  doi = {10.1103/PhysRevA.106.032607},
  url = {https://link.aps.org/doi/10.1103/PhysRevA.106.032607}
}

@article{subotnik_2018,
  title = {Universal approach to quantum thermodynamics in the strong coupling regime},
  author = {Dou, Wenjie and Ochoa, Maicol A. and Nitzan, Abraham and Subotnik, Joseph E.},
  journal = {Phys. Rev. B},
  volume = {98},
  issue = {13},
  pages = {134306},
  numpages = {15},
  year = {2018},
  month = {Oct},
  publisher = {American Physical Society},
  doi = {10.1103/PhysRevB.98.134306},
  url = {https://link.aps.org/doi/10.1103/PhysRevB.98.134306}
}

@article{Strasbegr_2019,
  title = {Repeated Interactions and Quantum Stochastic Thermodynamics at Strong Coupling},
  author = {Strasberg, Philipp},
  journal = {Phys. Rev. Lett.},
  volume = {123},
  issue = {18},
  pages = {180604},
  numpages = {7},
  year = {2019},
  month = {Oct},
  publisher = {American Physical Society},
  doi = {10.1103/PhysRevLett.123.180604},
  url = {https://link.aps.org/doi/10.1103/PhysRevLett.123.180604}
}

@article{Zhang_2021,
  title = {Non-Markovian decoherence dynamics of strong-coupling hybrid quantum systems: A master equation approach},
  author = {Chiang, Kai-Ting and Zhang, Wei-Min},
  journal = {Phys. Rev. A},
  volume = {103},
  issue = {1},
  pages = {013714},
  numpages = {18},
  year = {2021},
  month = {Jan},
  publisher = {American Physical Society},
  doi = {10.1103/PhysRevA.103.013714},
  url = {https://link.aps.org/doi/10.1103/PhysRevA.103.013714}
}

@article{Allahverdyan_2004,
doi = {10.1209/epl/i2004-10101-2},
url = {https://dx.doi.org/10.1209/epl/i2004-10101-2},
year = {2004},
month = {aug},
publisher = {},
volume = {67},
number = {4},
pages = {565},
author = {A. E. Allahverdyan and  R. Balian and  Th. M. Nieuwenhuizen},
title = {Maximal work extraction from finite quantum systems},
journal = {Europhysics Letters}
}

@article{Omkar2016,
  title = {The Unruh effect interpreted as a quantum noise channel},
  volume = {16},
  ISSN = {1533-7146},
  url = {http://dx.doi.org/10.26421/QIC16.9-10-2},
  DOI = {10.26421/qic16.9-10-2},
  number = {9 & 10},
  journal = {Quantum Information and Computation},
  publisher = {Rinton Press},
  author = {Omkar,  S. and Srikanth,  R. and Banerjee,  Subhashish and Alok,  Ashutosh Kumar},
  year = {2016},
  month = jul,
  pages = {757–770}
}

@article{Rivas_strong_coupling,
  title = {Strong Coupling Thermodynamics of Open Quantum Systems},
  author = {Rivas, \'Angel},
  journal = {Phys. Rev. Lett.},
  volume = {124},
  issue = {16},
  pages = {160601},
  numpages = {7},
  year = {2020},
  month = {Apr},
  publisher = {American Physical Society},
  doi = {10.1103/PhysRevLett.124.160601},
  url = {https://link.aps.org/doi/10.1103/PhysRevLett.124.160601}
}

@article{Strasberg_Esposito_2017,
  title = {Stochastic thermodynamics in the strong coupling regime: An unambiguous approach based on coarse graining},
  author = {Strasberg, Philipp and Esposito, Massimiliano},
  journal = {Phys. Rev. E},
  volume = {95},
  issue = {6},
  pages = {062101},
  numpages = {9},
  year = {2017},
  month = {Jun},
  publisher = {American Physical Society},
  doi = {10.1103/PhysRevE.95.062101},
  url = {https://link.aps.org/doi/10.1103/PhysRevE.95.062101}
}

@article{bose2025PT,
doi = {10.1088/1402-4896/ae64ba},
url = {https://doi.org/10.1088/1402-4896/ae64ba},
year = {2026},
month = {may},
publisher = {IOP Publishing},
volume = {101},
number = {18},
pages = {185103},
author = {Bose, Baibhab and Tiwari, Devvrat and Banerjee, Subhashish},
title = {{ \mathcal P }{ \mathcal T }-symmetric two-level open quantum systems: information theoretic facets},
journal = {Physica Scripta}
}

@Inbook{Miller2018_inbook,
author="Miller, Harry J. D.",
editor="Binder, Felix
and Correa, Luis A.
and Gogolin, Christian
and Anders, Janet
and Adesso, Gerardo",
title="Hamiltonian of Mean Force for Strongly-Coupled Systems",
bookTitle="Thermodynamics in the Quantum Regime: Fundamental Aspects and New Directions",
year="2018",
publisher="Springer International Publishing",
address="Cham",
pages="531--549",
isbn="978-3-319-99046-0",
doi="10.1007/978-3-319-99046-0_22",
url="https://doi.org/10.1007/978-3-319-99046-0_22"
}

@article{walk_Javid_2020,
  title = {Non-Markovian channel from the reduced dynamics of a coin in a quantum walk},
  author = {Naikoo, Javid and Banerjee, Subhashish and Chandrashekar, C. M.},
  journal = {Phys. Rev. A},
  volume = {102},
  issue = {6},
  pages = {062209},
  numpages = {9},
  year = {2020},
  month = {Dec},
  publisher = {American Physical Society},
  doi = {10.1103/PhysRevA.102.062209},
  url = {https://link.aps.org/doi/10.1103/PhysRevA.102.062209}
}

@article{ElGanainy2018,
  title = {Non-Hermitian physics and PT symmetry},
  volume = {14},
  ISSN = {1745-2481},
  url = {http://dx.doi.org/10.1038/nphys4323},
  DOI = {10.1038/nphys4323},
  number = {1},
  journal = {Nature Physics},
  publisher = {Springer Science and Business Media LLC},
  author = {El-Ganainy,  Ramy and Makris,  Konstantinos G. and Khajavikhan,  Mercedeh and Musslimani,  Ziad H. and Rotter,  Stefan and Christodoulides,  Demetrios N.},
  year = {2018},
  month = Jan,
  pages = {11–19}
}

@article{Feng2017,
  title = {Non-Hermitian photonics based on parity–time symmetry},
  volume = {11},
  ISSN = {1749-4893},
  url = {http://dx.doi.org/10.1038/s41566-017-0031-1},
  DOI = {10.1038/s41566-017-0031-1},
  number = {12},
  journal = {Nature Photonics},
  publisher = {Springer Science and Business Media LLC},
  author = {Feng,  Liang and El-Ganainy,  Ramy and Ge,  Li},
  year = {2017},
  month = Nov,
  pages = {752–762}
}

@article{Ashida2020,
  title = {Non-Hermitian physics},
  volume = {69},
  ISSN = {1460-6976},
  url = {http://dx.doi.org/10.1080/00018732.2021.1876991},
  DOI = {10.1080/00018732.2021.1876991},
  number = {3},
  journal = {Advances in Physics},
  publisher = {Informa UK Limited},
  author = {Ashida,  Yuto and Gong,  Zongping and Ueda,  Masahito},
  year = {2020},
  pages = {249–435}
}

@article{Bergholtz2021,
  title = {Exceptional topology of non-Hermitian systems},
  author = {Bergholtz, Emil J. and Budich, Jan Carl and Kunst, Flore K.},
  journal = {Rev. Mod. Phys.},
  volume = {93},
  issue = {1},
  pages = {015005},
  numpages = {31},
  year = {2021},
  month = {Feb},
  publisher = {American Physical Society},
  doi = {10.1103/RevModPhys.93.015005},
  url = {https://link.aps.org/doi/10.1103/RevModPhys.93.015005}
}

@article{Helbig2020,
  title = {Generalized bulk–boundary correspondence in non-Hermitian topolectrical circuits},
  volume = {16},
  ISSN = {1745-2481},
  url = {http://dx.doi.org/10.1038/s41567-020-0922-9},
  DOI = {10.1038/s41567-020-0922-9},
  number = {7},
  journal = {Nature Physics},
  publisher = {Springer Science and Business Media LLC},
  author = {Helbig,  T. and Hofmann,  T. and Imhof,  S. and Abdelghany,  M. and Kiessling,  T. and Molenkamp,  L. W. and Lee,  C. H. and Szameit,  A. and Greiter,  M. and Thomale,  R.},
  year = {2020},
  pages = {747–750}
}

@article{Hofmann2020,
  title = {Reciprocal skin effect and its realization in a topolectrical circuit},
  author = {Hofmann, Tobias and Helbig, Tobias and Schindler, Frank and Salgo, Nora and Brzezi\ifmmode \acute{n}\else \'{n}\fi{}ska, Marta and Greiter, Martin and Kiessling, Tobias and Wolf, David and Vollhardt, Achim and Kaba\ifmmode \check{s}\else \v{s}\fi{}i, Anton and Lee, Ching Hua and Bilu\ifmmode \check{s}\else \v{s}\fi{}i\ifmmode \acute{c}\else \'{c}\fi{}, Ante and Thomale, Ronny and Neupert, Titus},
  journal = {Phys. Rev. Res.},
  volume = {2},
  issue = {2},
  pages = {023265},
  numpages = {11},
  year = {2020},
  month = {Jun},
  publisher = {American Physical Society},
  doi = {10.1103/PhysRevResearch.2.023265},
  url = {https://link.aps.org/doi/10.1103/PhysRevResearch.2.023265}
}

@article{Li2019,
  title = {Observation of parity-time symmetry breaking transitions in a dissipative Floquet system of ultracold atoms},
  volume = {10},
  ISSN = {2041-1723},
  url = {http://dx.doi.org/10.1038/s41467-019-08596-1},
  DOI = {10.1038/s41467-019-08596-1},
  number = {1},
  journal = {Nature Communications},
  publisher = {Springer Science and Business Media LLC},
  author = {Li,  Jiaming and Harter,  Andrew K. and Liu,  Ji and de Melo,  Leonardo and Joglekar,  Yogesh N. and Luo,  Le},
  year = {2019},
  month = Feb 
}

@article{Lapp2019,
  title = {Engineering tunable local loss in a synthetic lattice of momentum states},
  volume = {21},
  ISSN = {1367-2630},
  url = {http://dx.doi.org/10.1088/1367-2630/ab1147},
  DOI = {10.1088/1367-2630/ab1147},
  number = {4},
  journal = {New Journal of Physics},
  publisher = {IOP Publishing},
  author = {Lapp,  Samantha and Ang’ong’a,  Jackson and An,  Fangzhao Alex and Gadway,  Bryce},
  year = {2019},
  month = Apr,
  pages = {045006}
}

@book{Moiseyev2011,
  title={Non-Hermitian Quantum Mechanics},
  author={Moiseyev, N.},
  isbn={9781139496995},
  url={https://books.google.co.in/books?id=QwAXVvk_57QC},
  year={2011},
  publisher={Cambridge University Press}
}

@article{Rotter2009,
  title = {A non-Hermitian Hamilton operator and the physics of open quantum systems},
  volume = {42},
  ISSN = {1751-8121},
  url = {http://dx.doi.org/10.1088/1751-8113/42/15/153001},
  DOI = {10.1088/1751-8113/42/15/153001},
  number = {15},
  journal = {Journal of Physics A: Mathematical and Theoretical},
  publisher = {IOP Publishing},
  author = {Rotter,  Ingrid},
  year = {2009},
  month = Mar,
  pages = {153001}
}

@article{Bender2012,
  title = {Ordinary versus $\mathcal{P}\mathcal{T}$-symmetric ${\ensuremath{\phi}}^{3}$ quantum field theory},
  author = {Bender, Carl M. and Branchina, Vincenzo and Messina, Emanuele},
  journal = {Phys. Rev. D},
  volume = {85},
  issue = {8},
  pages = {085001},
  numpages = {7},
  year = {2012},
  month = {Apr},
  publisher = {American Physical Society},
  doi = {10.1103/PhysRevD.85.085001},
  url = {https://link.aps.org/doi/10.1103/PhysRevD.85.085001}
}

@article{Bender2018fieldtheory,
  title = {$\mathcal{P}\mathcal{T}$-symmetric quantum field theory in $D$ dimensions},
  author = {Bender, Carl M. and Hassanpour, Nima and Klevansky, S. P. and Sarkar, Sarben},
  journal = {Phys. Rev. D},
  volume = {98},
  issue = {12},
  pages = {125003},
  numpages = {9},
  year = {2018},
  month = {Dec},
  publisher = {American Physical Society},
  doi = {10.1103/PhysRevD.98.125003},
  url = {https://link.aps.org/doi/10.1103/PhysRevD.98.125003}
}

@article{Brachio1,
  title = {$\mathcal{P}\mathcal{T}$-symmetric brachistochrone problem, Lorentz boosts, and nonunitary operator equivalence classes},
  author = {G\"unther, Uwe and Samsonov, Boris F.},
  journal = {Phys. Rev. A},
  volume = {78},
  issue = {4},
  pages = {042115},
  numpages = {9},
  year = {2008},
  month = {Oct},
  publisher = {American Physical Society},
  doi = {10.1103/PhysRevA.78.042115},
  url = {https://link.aps.org/doi/10.1103/PhysRevA.78.042115}
}

@article{Brachio2,
  title = {Naimark-Dilated $\mathcal{P}\mathcal{T}$-Symmetric Brachistochrone},
  author = {G\"unther, Uwe and Samsonov, Boris F.},
  journal = {Phys. Rev. Lett.},
  volume = {101},
  issue = {23},
  pages = {230404},
  numpages = {4},
  year = {2008},
  month = {Dec},
  publisher = {American Physical Society},
  doi = {10.1103/PhysRevLett.101.230404},
  url = {https://link.aps.org/doi/10.1103/PhysRevLett.101.230404}
}

@article{Zheng2013,
    author = {Zheng, Chao and Hao, Liang and Long, Gui Lu},
    title = {Observation of a fast evolution in a parity-time-symmetric system},
    journal = {Philosophical Transactions of the Royal Society A: Mathematical, Physical and Engineering Sciences},
    volume = {371},
    number = {1989},
    pages = {20120053},
    year = {2013},
    month = {04},
    issn = {1364-503X},
    doi = {10.1098/rsta.2012.0053},
    url = {https://doi.org/10.1098/rsta.2012.0053},
    eprint = {https://royalsocietypublishing.org/rsta/article-pdf/doi/10.1098/rsta.2012.0053/419738/rsta.2012.0053.pdf},
}

@article{CrokePT_information,
  title = {$\mathcal{PT}$-symmetric Hamiltonians and their application in quantum information},
  author = {Croke, Sarah},
  journal = {Phys. Rev. A},
  volume = {91},
  issue = {5},
  pages = {052113},
  numpages = {12},
  year = {2015},
  month = {May},
  publisher = {American Physical Society},
  doi = {10.1103/PhysRevA.91.052113},
  url = {https://link.aps.org/doi/10.1103/PhysRevA.91.052113}
}

@article{Brody2016,
  title = {Consistency of PT-symmetric quantum mechanics},
  volume = {49},
  ISSN = {1751-8121},
  url = {http://dx.doi.org/10.1088/1751-8113/49/10/10LT03},
  DOI = {10.1088/1751-8113/49/10/10lt03},
  number = {10},
  journal = {Journal of Physics A: Mathematical and Theoretical},
  publisher = {IOP Publishing},
  author = {Brody,  Dorje C},
  year = {2016},
  month = Jan,
  pages = {10LT03}
}

@article{Brody2013,
  title = {Biorthogonal quantum mechanics},
  volume = {47},
  ISSN = {1751-8121},
  url = {http://dx.doi.org/10.1088/1751-8113/47/3/035305},
  DOI = {10.1088/1751-8113/47/3/035305},
  number = {3},
  journal = {Journal of Physics A: Mathematical and Theoretical},
  publisher = {IOP Publishing},
  author = {Brody,  Dorje C},
  year = {2013},
  month = Dec,
  pages = {035305}
}

@article{Deffner2015,
  title = {Jarzynski Equality in $\mathcal{P}\mathcal{T}$-Symmetric Quantum Mechanics},
  author = {Deffner, Sebastian and Saxena, Avadh},
  journal = {Phys. Rev. Lett.},
  volume = {114},
  issue = {15},
  pages = {150601},
  numpages = {5},
  year = {2015},
  month = {Apr},
  publisher = {American Physical Society},
  doi = {10.1103/PhysRevLett.114.150601},
  url = {https://link.aps.org/doi/10.1103/PhysRevLett.114.150601}
}

@article{Peng2014NatureComm,
  author = {Peng, B. et al.},
  title = {Parity–time-symmetric whispering-gallery microcavities},
  journal = {Nature Communications},
  volume = {5},
  pages = {15791},
  year = {2014},
  doi = {10.1038/ncomms15791}
}

@article{Xiao2017,
  title = {Observation of topological edge states in parity–time-symmetric quantum walks},
  volume = {13},
  ISSN = {1745-2481},
  url = {http://dx.doi.org/10.1038/nphys4204},
  DOI = {10.1038/nphys4204},
  number = {11},
  journal = {Nature Physics},
  publisher = {Springer Science and Business Media LLC},
  author = {Xiao,  L. and Zhan,  X. and Bian,  Z. H. and Wang,  K. K. and Zhang,  X. and Wang,  X. P. and Li,  J. and Mochizuki,  K. and Kim,  D. and Kawakami,  N. and Yi,  W. and Obuse,  H. and Sanders,  B. C. and Xue,  P.},
  year = {2017},
  pages = {1117–1123}
}

@article{Wen2019,
  title = {Experimental demonstration of a digital quantum simulation of a general $\mathcal{PT}$-symmetric system},
  author = {Wen, Jingwei and Zheng, Chao and Kong, Xiangyu and Wei, Shijie and Xin, Tao and Long, Guilu},
  journal = {Phys. Rev. A},
  volume = {99},
  issue = {6},
  pages = {062122},
  numpages = {8},
  year = {2019},
  month = {Jun},
  publisher = {American Physical Society},
  doi = {10.1103/PhysRevA.99.062122},
  url = {https://link.aps.org/doi/10.1103/PhysRevA.99.062122}
}

@article{Wang2020,
  title = {Experimental Investigation of State Distinguishability in Parity-Time Symmetric Quantum Dynamics},
  author = {Wang, Yi-Tao and Li, Zhi-Peng and Yu, Shang and Ke, Zhi-Jin and Liu, Wei and Meng, Yu and Yang, Yuan-Ze and Tang, Jian-Shun and Li, Chuan-Feng and Guo, Guang-Can},
  journal = {Phys. Rev. Lett.},
  volume = {124},
  issue = {23},
  pages = {230402},
  numpages = {7},
  year = {2020},
  month = {Jun},
  publisher = {American Physical Society},
  doi = {10.1103/PhysRevLett.124.230402},
  url = {https://link.aps.org/doi/10.1103/PhysRevLett.124.230402}
}

@article{Choutri2017,
  title = {Pseudo-Hermitian Systems with P T $\mathcal {P}\mathcal {T}$ -Symmetry: Degeneracy and Krein Space},
  volume = {56},
  ISSN = {1572-9575},
  url = {http://dx.doi.org/10.1007/s10773-017-3299-5},
  DOI = {10.1007/s10773-017-3299-5},
  number = {5},
  journal = {International Journal of Theoretical Physics},
  publisher = {Springer Science and Business Media LLC},
  author = {Choutri,  B. and Cherbal,  O. and Ighezou,  F. Z. and Drir,  M.},
  year = {2017},
  month = Feb,
  pages = {1595–1604}
}

@article{Gao:21,
author = {Wei-Chao Gao and Chao Zheng and Lu Liu and Tie-Jun Wang and Chuan Wang},
journal = {Opt. Express},
number = {1},
pages = {517--526},
publisher = {Optica Publishing Group},
title = {Experimental simulation of the parity-time symmetric dynamics using photonic qubits},
volume = {29},
month = {Jan},
year = {2021},
url = {https://opg.optica.org/oe/abstract.cfm?URI=oe-29-1-517},
doi = {10.1364/OE.405815},
}

@Article{PTopen1,
	title={{Emergence of PT-symmetry breaking in open quantum systems}},
	author={Julian Huber and Peter Kirton and Stefan Rotter and Peter Rabl},
	journal={SciPost Phys.},
	volume={9},
	pages={052},
	year={2020},
	publisher={SciPost},
	doi={10.21468/SciPostPhys.9.4.052},
	url={https://scipost.org/10.21468/SciPostPhys.9.4.052},
}

@article{PTopen2Brody_2021,
doi = {10.1088/1742-6596/2038/1/012005},
url = {https://doi.org/10.1088/1742-6596/2038/1/012005},
year = {2021},
month = {oct},
publisher = {IOP Publishing},
volume = {2038},
number = {1},
pages = {012005},
author = {Brody, Dorje C},
title = {PT symmetry and the evolution speed in open quantum systems 1},
journal = {Journal of Physics: Conference Series}
}

@article{Ashida2017,
  title = {Parity-time-symmetric quantum critical phenomena},
  volume = {8},
  ISSN = {2041-1723},
  url = {http://dx.doi.org/10.1038/ncomms15791},
  DOI = {10.1038/ncomms15791},
  number = {1},
  journal = {Nature Communications},
  publisher = {Springer Science and Business Media LLC},
  author = {Ashida,  Yuto and Furukawa,  Shunsuke and Ueda,  Masahito},
  year = {2017} 
}

@article{Stenholm01042004,
author = {Stig Stenholm and Matthias Jakob},
title = {Open systems and time reversal},
journal = {Journal of Modern Optics},
volume = {51},
number = {6-7},
pages = {841--850},
year = {2004},
publisher = {Taylor \& Francis},
doi = {10.1080/09500340408233601},
URL = {https://www.tandfonline.com/doi/abs/10.1080/09500340408233601}
}

@article{PTopen3,
  title = {$\mathcal{PT}$ phase transition in open quantum systems with Lindblad dynamics},
  author = {Nakanishi, Yuma and Sasamoto, Tomohiro},
  journal = {Phys. Rev. A},
  volume = {105},
  issue = {2},
  pages = {022219},
  numpages = {17},
  year = {2022},
  month = {Feb},
  publisher = {American Physical Society},
  doi = {10.1103/PhysRevA.105.022219},
  url = {https://link.aps.org/doi/10.1103/PhysRevA.105.022219}
}

@article{PRAVEENA2023171260,
title = {A review: Rise of PT-symmetry for laser applications},
journal = {Optik},
volume = {289},
pages = {171260},
year = {2023},
issn = {0030-4026},
doi = {https://doi.org/10.1016/j.ijleo.2023.171260},
url = {https://www.sciencedirect.com/science/article/pii/S003040262300757X},
author = {S. Praveena and K. Senthilnathan}
}

@article{Fluctu1,
  title = {Finite bath fluctuation theorem},
  author = {Campisi, Michele and Talkner, Peter and H\"anggi, Peter},
  journal = {Phys. Rev. E},
  volume = {80},
  issue = {3},
  pages = {031145},
  numpages = {9},
  year = {2009},
  month = {Sep},
  publisher = {American Physical Society},
  doi = {10.1103/PhysRevE.80.031145},
  url = {https://link.aps.org/doi/10.1103/PhysRevE.80.031145}
}

@article{Fluctu2,
  title = {Colloquium: Quantum fluctuation relations: Foundations and applications},
  author = {Campisi, Michele and H\"anggi, Peter and Talkner, Peter},
  journal = {Rev. Mod. Phys.},
  volume = {83},
  issue = {3},
  pages = {771--791},
  numpages = {0},
  year = {2011},
  month = {Jul},
  publisher = {American Physical Society},
  doi = {10.1103/RevModPhys.83.771},
  url = {https://link.aps.org/doi/10.1103/RevModPhys.83.771}
}

@article{exceptional1,
  title = {Parity–time symmetry and exceptional points in photonics},
  volume = {18},
  ISSN = {1476-4660},
  url = {http://dx.doi.org/10.1038/s41563-019-0304-9},
  DOI = {10.1038/s41563-019-0304-9},
  number = {8},
  journal = {Nature Materials},
  publisher = {Springer Science and Business Media LLC},
  author = {\"{O}zdemir,  \c{S}. K. and Rotter,  S. and Nori,  F. and Yang,  L.},
  year = {2019},
  month = Apr,
  pages = {783–798}
}

@article{exceptional2,
  title = {Exceptional-point-locked PT-symmetric system for ultrasensitive displacement sensing via amplitude readout},
  volume = {128},
  ISSN = {1077-3118},
  url = {http://dx.doi.org/10.1063/5.0296728},
  DOI = {10.1063/5.0296728},
  number = {1},
  journal = {Applied Physics Letters},
  publisher = {AIP Publishing},
  author = {Yin,  Ke and Hao,  Xianglin and Tang,  Kaihao and Tan,  Lu and Dong,  Tianyu and Zhu,  Huacheng and Yang,  Yang},
  year = {2026},
  month = Jan 
}

@article{exceptional3,
  title = {Topological exceptional surfaces in non-Hermitian systems with parity-time and parity-particle-hole symmetries},
  author = {Okugawa, Ryo and Yokoyama, Takehito},
  journal = {Phys. Rev. B},
  volume = {99},
  issue = {4},
  pages = {041202},
  numpages = {6},
  year = {2019},
  month = {Jan},
  publisher = {American Physical Society},
  doi = {10.1103/PhysRevB.99.041202},
  url = {https://link.aps.org/doi/10.1103/PhysRevB.99.041202}
}

@article{exceptional4,
  title = {Symmetry-Protected Multifold Exceptional Points and Their Topological Characterization},
  author = {Delplace, Pierre and Yoshida, Tsuneya and Hatsugai, Yasuhiro},
  journal = {Phys. Rev. Lett.},
  volume = {127},
  issue = {18},
  pages = {186602},
  numpages = {6},
  year = {2021},
  month = {Oct},
  publisher = {American Physical Society},
  doi = {10.1103/PhysRevLett.127.186602},
  url = {https://link.aps.org/doi/10.1103/PhysRevLett.127.186602}
}

@article{exceptional0,
  title = {The physics of exceptional points},
  volume = {45},
  ISSN = {1751-8121},
  url = {http://dx.doi.org/10.1088/1751-8113/45/44/444016},
  DOI = {10.1088/1751-8113/45/44/444016},
  number = {44},
  journal = {Journal of Physics A: Mathematical and Theoretical},
  publisher = {IOP Publishing},
  author = {Heiss,  W D},
  year = {2012},
  month = Oct,
  pages = {444016}
}

@article{SimilarityinPT,
    author = {Montag, Anton and Kunst, Flore K.},
    title = {Essential implications of similarities in non-Hermitian systems},
    journal = {Journal of Mathematical Physics},
    volume = {65},
    number = {12},
    pages = {122101},
    year = {2024},
    month = {12},
    issn = {0022-2488},
    doi = {10.1063/5.0206211},
    url = {https://doi.org/10.1063/5.0206211}
}

@article{SimilarityAshida02072020,
author = {Yuto Ashida and Zongping Gong and Masahito Ueda},
title = {Non-Hermitian physics},
journal = {Advances in Physics},
volume = {69},
number = {3},
pages = {249--435},
year = {2020},
publisher = {Taylor \& Francis},
doi = {10.1080/00018732.2021.1876991},


URL = { 
    
        https://doi.org/10.1080/00018732.2021.1876991
    
    

},
eprint = { 
    
        https://doi.org/10.1080/00018732.2021.1876991
    
    

}

}

@article{similarity3,
    author = {Zhang, Ruili and Qin, Hong and Xiao, Jianyuan},
    title = {PT-symmetry entails pseudo-Hermiticity regardless of diagonalizability},
    journal = {Journal of Mathematical Physics},
    volume = {61},
    number = {1},
    pages = {012101},
    year = {2020},
    month = {01},
    issn = {0022-2488},
    doi = {10.1063/1.5117211},
    url = {https://doi.org/10.1063/1.5117211},
}

@article{similarity2Mostafazadeh2003,
  title = {Exact<i>PT</i>-symmetry is equivalent to Hermiticity},
  volume = {36},
  ISSN = {1361-6447},
  url = {http://dx.doi.org/10.1088/0305-4470/36/25/312},
  DOI = {10.1088/0305-4470/36/25/312},
  number = {25},
  journal = {Journal of Physics A: Mathematical and General},
  publisher = {IOP Publishing},
  author = {Mostafazadeh,  Ali},
  year = {2003},
  pages = {7081–7091}
}

@misc{kleefeld2009constructiongeneralinnerproduct,
      title={The construction of a general inner product in non-Hermitian quantum theory and some explanation for the nonuniqueness of the C operator in PT quantum mechanics}, 
      author={F. Kleefeld},
      year={2009},
      eprint={0906.1011},
      archivePrefix={arXiv},
      primaryClass={hep-th},
      url={https://arxiv.org/abs/0906.1011}, 
}

@misc{kleefeld2023equivalencepaisuhlenbeckoscillatormodel,
      title={On the equivalence of the Pais-Uhlenbeck oscillator model and two non-Hermitian Harmonic Oscillators}, 
      author={Frieder Kleefeld},
      year={2023},
      eprint={2302.14621},
      archivePrefix={arXiv},
      primaryClass={quant-ph},
      url={https://arxiv.org/abs/2302.14621}, 
}

@ARTICLE{jaynes_cummings,
  author={Jaynes, E.T. and Cummings, F.W.},
  journal={Proceedings of the IEEE}, 
  title={Comparison of quantum and semiclassical radiation theories with application to the beam maser}, 
  year={1963},
  volume={51},
  number={1},
  pages={89-109},
  doi={10.1109/PROC.1963.1664}}

\end{document}